\documentclass[10pt,a4paper]{article}
\pdfoutput=1
\usepackage{jheppub}
\usepackage{epsfig}
\usepackage{graphicx}
\usepackage{multirow}
\usepackage{amsmath}
\usepackage{amssymb}
\usepackage{slashed}
\usepackage{slashed}
\usepackage[dvipsnames]{xcolor}

\DeclareMathAlphabet{\mathpzc}{OT1}{pzc}{m}{it}
\makeatletter
\newbox\slashbox \setbox\slashbox=\hbox{$/$}
\newbox\Slashbox \setbox\Slashbox=\hbox{\large$/$}
\def\pFMslash#1{\setbox\@tempboxa=\hbox{$#1$}
  \@tempdima=0.5\wd\slashbox \advance\@tempdima 0.5\wd\@tempboxa
  \copy\slashbox \kern-\@tempdima \box\@tempboxa}
\def\pFMSlash#1{\setbox\@tempboxa=\hbox{$#1$}
  \@tempdima=0.5\wd\Slashbox \advance\@tempdima 0.5\wd\@tempboxa
  \copy\Slashbox \kern-\@tempdima \box\@tempboxa}

\def\miss#1{\ifmmode{/\mkern-11mu #1}\else{${/\mkern-11mu #1}$}\fi}
\makeatother

\title{Flavor violating leptonic decays of $\tau$ and $\mu$ leptons in the Standard Model with massive neutrinos}

\author[1]{G. Hern\'andez-Tom\'e}
\author[1]{, G. L\'opez Castro}
\author[1]{and P. Roig}

 \affiliation[1]{Departamento de F\'isica, Centro de Investigaci\'on y de Estudios Avanzados del Instituto Polit\'ecnico Nacional,
Apdo. Postal 14-740, 07000 M\'exico D.F., M\'exico}

\emailAdd{ghernandez@fis.cinvestav.mx}
\emailAdd{glopez@fis.cinvestav.mx}
\emailAdd{proig@fis.cinvestav.mx}

\abstract{We have revisited the computations of the flavor violating leptonic decays of the $\tau$ and $\mu$ leptons into three lighter charged leptons in the Standard Model with non-vanishing neutrino masses. We were driven by a claimed unnaturally large branching ratio predicted for the $\tau^-\to \mu^- \ell^+ \ell^-$ ($\ell=\mu, e$) decays \cite{Pham:1998fq}, which was at odds with the corresponding predictions for the $\mu^-\to e^- e^-e^+$ processes \cite{Petcov:1976ff}. In contrast with the prediction in \cite{Pham:1998fq}, our results are strongly suppressed and in good agreement with the approximation done in ref.~\cite{Petcov:1976ff}, where masses and momenta of the external particles were neglected in order to deal with the loop integrals. However, as a result of keeping external momenta and masses in the computation of the dominant penguin and box diagrams- we even find slightly smaller branching fractions. Therefore, we confirm that any future observation of such processes would be an unambiguous manifestation of new physics beyond the Standard Model.

}

\begin{document}

\maketitle
\flushbottom

\section{Introduction}\label{Intro}

Lepton flavor violating (LFV) processes are forbidden in the standard model (SM) \cite{SM} with massless neutrinos. 
However, the experimental evidence of neutrino oscillations \cite{nuosc} claims for an extended model with neutrino mass terms. 
For massive neutrinos, the mass matrix will be nondiagonal in the interaction (weak) basis, as occurs in the quark sector \cite{CKM}, and the mixing of three light neutrinos could be described through the $3\times 3$ unitary Pontecorvo-Maki-Nakagawa-Sakata (PMNS) matrix \cite{PMNS}. In such scenario, charged LFV transitions could arise, for instance, from one loop diagrams involving a couple of $W\ell\nu_\ell$ vertices with different flavor neutrinos each. However, it turns out natural having a strong suppression for this class of processes owing to a GIM-like mechanism \cite{Glashow:1970gm}, just as it has been reported for the $\mu^-\to e^- \gamma$ decay, with a prediction at an unobservable low rate: $BR(\mu^-\to e^-\gamma)\sim \mathcal{O}(10^{-55})$ \cite{Petcov:1976ff, LibroCheng, Calibbi:2017uvl}, which is far away from the capacity of any current or foreseen experimental facility.

\begin{figure}[h!]
\begin{center}
    \includegraphics[scale=0.7]{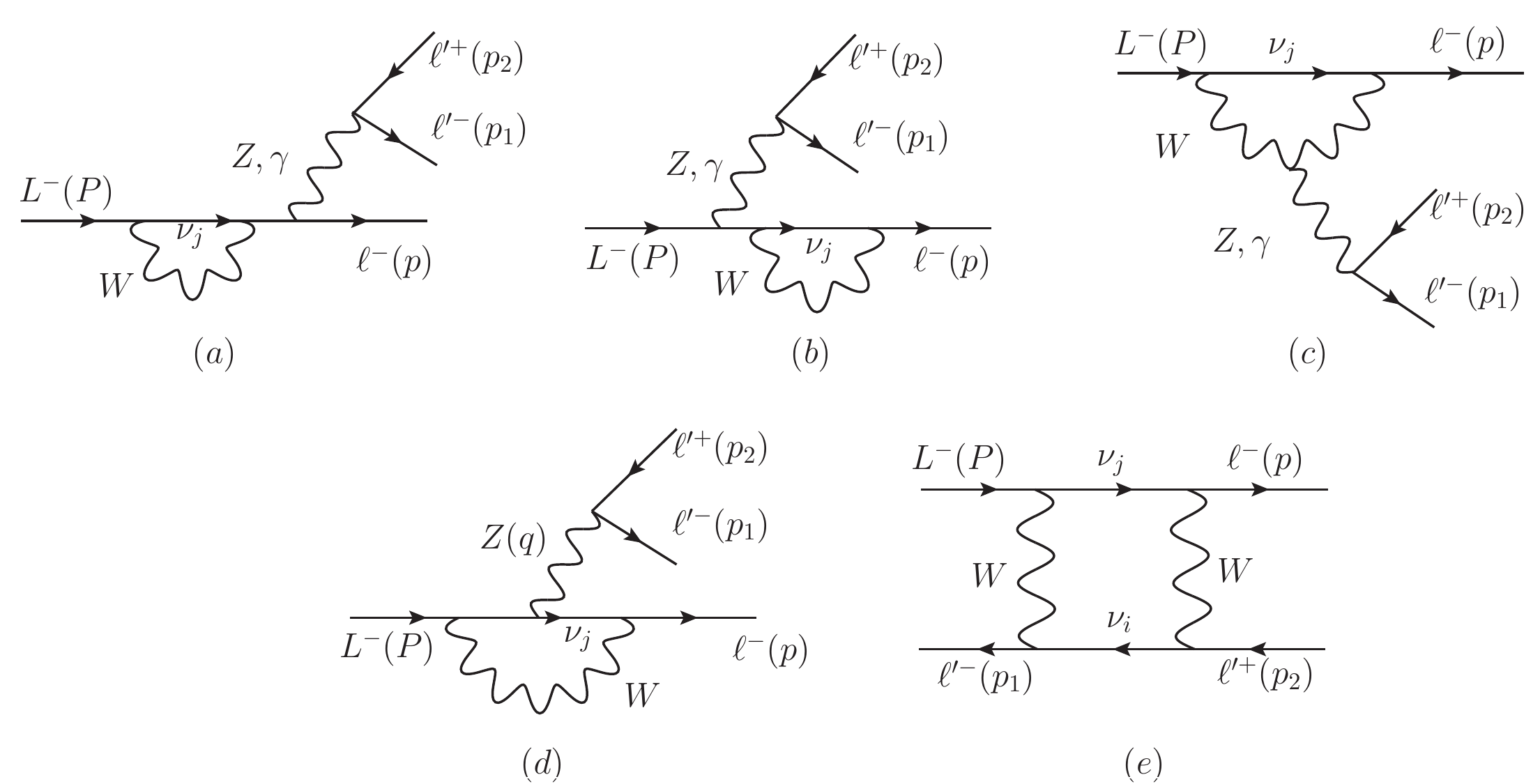}
    \caption{Feynman diagrams for the $L^{-}\to \ell^- \ell^{\prime -} \ell^{\prime +}$ decays, in the presence of lepton mixing (\textit{i. e.}, 
    non-vanishing neutrino masses). In `renormalizable' $R_\xi$ gauges, similar diagrams need to be added, which are obtained replacing the $W$ gauge bosons by the respective would-be Goldstone bosons. Notice that diagram (d) only involves the $Z$ gauge boson, whereas the (a), (b) and (c) diagrams can also be mediated by the photon. Additionally, when  $\ell=\ell^{\prime}$ similar contributions (exchanging $p \leftrightarrow p_1$) to the amplitudes of diagrams (a) to (e) must be subtracted in order to antisymmetrize the amplitude. On the other hand, when $\ell\neq\ell^\prime$, owing to the fact that the neutral gauge bosons $\gamma$ and $Z$ do not change flavor, only a similar (e) box diagram must be added interchanging $\ell(p)\leftrightarrow\ell^{\prime}(p_1)$.   }
     \label{diagrams}
         \end{center} 
\end{figure}

By way of contrast, the prediction for the $\tau^-\to \mu^- \ell^+ \ell^-$ ($\ell=\mu, e$) decays given by ref.~\cite{Pham:1998fq} indicates that the GIM cancellation for these processes is much milder and a value of $BR(\tau^- \to \mu^- \ell^+\ell^-)\geq 10^{-14}$ is reported. An updated evaluation using the amplitude derived in ref.~\cite{Pham:1998fq}, employing the latest global fit results for neutrino mixing \cite{Patrignani:2016xqp, Fits} yields a branching fraction $\sim4\cdot 10^{-16}$ for the three muon channel. Both values are still far away from the PDG upper bounds, $1.5\cdot10^{-8}$ (for $\ell=e$) and $2.1\cdot10^{-8}$ ($\ell=\mu$) at $90\%$ confidence level~\footnote{More stringent bounds  of $1.1 \cdot 10^{-8}$ and $1.2 \cdot 10^{-8}$, respectively, can be  obtained by combining results of different experiments according to the  HFLAV group \cite{Referencia}. Belle-II shall be able to set limits on the $\tau^-\to\mu^-\mu^+\mu^-$ decay at the level of $3\cdot10^{-10}$ with their full data set ($50$ ab$^{-1}$) \cite{Kou:2018nap}.}. Similarly, we verified that using the values reported in refs.~\cite{Patrignani:2016xqp, Fits} for the neutrino mixing parameters, Pham's result \cite{Pham:1998fq} would predict a $\mu^-\to e^-e^+e^-$ branching ratio of $\sim 2\cdot10^{-21}$, larger than Petcov's prediction ($\sim 10^{-53}$ evaluated with updated neutrino masses and mixings input)  \cite{Petcov:1976ff} by at least some thirty orders of magnitude. Again, the current upper limit on this decay channel ($1\cdot10^{-12}$ at $90\%$ C.L. \cite{Patrignani:2016xqp}) is still far from testing Pham's result \cite{Pham:1998fq}. This author claims that this unexpectedly large estimation is due to the presence of a divergent logarithmic term depending on the neutrino mass, which comes from a one-loop diagram that involves two neutrino propagators (diag. (d) in our fig. \ref{diagrams}).

Certainly, considering effects or processes that arise from quantum corrections could involve divergent loop integrals. However, in any renormalizable theory, the possible divergences must vanish order by order (in the loop or effective field theory expansion) to be able to  define (finite) observables. In fact, in a QFT the divergences can be classified into two types: ultraviolet (UV) and infrared divergences (IR). The former (UV) appear in the high-energy regime and they can be healed redefining the theory parameters, whereas the latter (IR) occur in the low-energy regime and can be classified in soft and collinear divergences, which cancel however in properly defined (\textit{IR-safe}) observables \cite{KLN}. We show that the seeming logarithmic divergent behavior of the LFV amplitude  reported in ref. \cite{Pham:1998fq} is not present, as the vanishing momentum transfer approximation considered in that paper lies outside the physical region. Consequently, the rates of $L^{-}\to \ell^- \ell^{\prime -} \ell^{\prime +}$ decays in the SM extended with massive neutrinos are extremely suppressed, in agreement with ref.~\cite{Petcov:1976ff}. It is worth noting that the LFV amplitudes must vanish in the limit of massless neutrinos. This requirement is satisfied by the result of Ref. [8], but it is not the case in ref, [10] which behaves as $\sum_j U_{Lj}U_{\ell j}^*
\log(m_j/m_W)$ for very small neutrino masses. Our result, as it will be shown below, satisfies the
expected agreement with the SM.

In section 2 and 3 we discuss in detail our computation of these processes and compare it to those in refs. \cite{Petcov:1976ff} and \cite{Pham:1998fq}, showing explicitly why the approximation in \cite{Pham:1998fq} is unreliable, and reproducing the results of \cite{Petcov:1976ff} in the approximation where masses and momenta of the external particles are neglected from the beginning. However, we also analize
the numerical accuracy of this approximation. Finally, we state our conclusions in section 4. Several appendices complete technical details of our calculation.
\\\\

\section{Z-Penguin contribution emission from internal neutrino line}\label{Comput}

The $L^{-}\to \ell^- \ell^{\prime -} \ell^{\prime +}$ decays can be induced through the diagrams depicted in fig. \ref{diagrams}. 
Since the main purpose of this work is to falsify the existence of the logarithmic divergent term claimed in ref.~\cite{Pham:1998fq}, we first concentrate on the amplitude of the diagram (d). We have, however, verified the corresponding expressions for the loop integrals in ref.~\cite{Petcov:1976ff} for the particular process $\mu^-\to e^- e^-e^+$, when masses and momenta of external leptons are neglected in the computations. Particularly, in Ref. \cite{Petcov:1976ff} it is shown that the corresponding branching ratio is completely dominated by those diagrams with two neutrino propagators, \textit{i. e.} (d) and (e) in fig. \ref{diagrams}, which contribute comparably. 

In our analysis, we keep employing the convention used by ref.~\cite{Pham:1998fq}, in order to denote the masses and momenta (see fig. \ref{diagrams}) of the external leptons, that is $M$ and $P$ ($m$ and $p$) stand for the mass and momentum of the $L^-$ ($\ell^-$) lepton, respectively. In this way, the amplitude of the diagram (d) can be written as

\begin{eqnarray}
\mathcal{M}_{d}&\sim &\frac{i}{m_Z^2}l_{L\ell}^{\lambda}\times \ell_{\ell'\ell'\lambda},
\end{eqnarray}
where $\ell_{\ell'\ell'}^\lambda=-ig/(2c_W)\bar{u}(p_{1})\gamma^{\lambda}(g_v^{\ell'}-g_a^{\ell'} \gamma_5) v(p_2)$ 
\footnote{$g$ is the $SU(2)_L$ coupling and $c_W(s_W)$ is short for the cosine(sine) of the weak mixing angle $\theta_W$. In the SM, $g_v^{\ell'}=-1/2+2 s_W^2$ and $g_a^{\ell'}=-1/2$.} is independent of the loop integration, whereas the relevant part for the latter is given by the effective $ZL\ell$ transition as follows:

\begin{equation}
l_{L\ell}^{\lambda}=\left(\frac{-ig}{4c_W} \right)\left(\frac{-ig}{2\sqrt{2}} \right)^2\sum_{j=1}^3 U^*_{\ell j}U_{Lj}\bar{u}(p)\Gamma^\lambda_{j} u(P),\label{parteloop}
\end{equation}
where $U_{im}$ are entries of the PMNS mixing matrix. In the Feynman-'t Hooft gauge we have

\begin{equation}
\Gamma^\lambda_{j}=\int \frac{d^4 k}{(2\pi)^4}\frac{\gamma_{\rho}(1-\gamma_5)i\left[(\slashed{p}+\slashed{k})+m_j \right]\gamma^{\lambda}(1-\gamma_5)i\left[(\slashed{P}+\slashed{k})+m_j \right]\gamma_{\sigma}(1-\gamma_5)(-ig^{\rho\sigma})}{\left[(p+k)^2-m_j^2 \right]\left[(P+k)^2-m_j^2 \right]\left[k^2-m_W^2 \right]}.\label{loop-int-penguin}
\end{equation}
After making the loop integration using dimensional regularization in order to deal with the (logarithmic) UV divergences, the Lorentz structure for the $\Gamma^\lambda_{j}$ factor can be written as follows, 
 
\begin{eqnarray}
\Gamma^\lambda_{j}&=&F_a\gamma^{\lambda}(1-\gamma_5)+F_b\gamma^{\lambda}(1+\gamma_5)+F_c(P+p)^{\lambda}(1+\gamma_5)\nonumber\\ &+&F_d(P+p)^{\lambda}(1-\gamma_5)+F_e q^{\lambda}(1+\gamma_5)+F_f q^{\lambda}(1-\gamma_5),
\end{eqnarray}
where in general $F_k=F_k(q^2,m_j^2)$  $(k=a,\,b...,\,f)$ are functions given in terms of the momentum transfer $q^2$, and the neutrino mass squared (of course $F_k$ functions will also depend on the mass of the $W$ gauge boson and external masses, but these have well-defined values).

At this point, it is worth to note that in the approximation where the momenta of the external particles are neglected in equation (\ref{loop-int-penguin}), such as it is done in ref.~\cite{Petcov:1976ff} for the $\mu\to3e$ decay, the computation is simplified considerably, as the only possible contribution is given by the $F^0_a$ function, where we are using a superscript $0$ in order to distinguish this approximation. 
In this simple case, the $F^0_{a}$ function will not depend on $q^2$ and is given in terms of the Feynman parameters as follows 

\begin{equation}
F^{0}_{a}(m_j^2)=\frac{1}{2\pi^2}\int_0^1\int_0^{1-x}\left[2 + \log \left(D^0_j/\mu^2\right)\right]dxdy,\label{Petcov-penguin-Feynman}
\end{equation}
where  $D^0_j(m_j^2)=(1-x) m_j^2+x m_W^2$. Whereas in terms of PaVe functions it is given by 

\begin{eqnarray}
F^0_{a}(m_j^2)&=&-\frac{1}{8\pi^2 \left(m_j^2-m_W^2\right){}^2}\left[ 2 m_j^2 \left(m_j^2-2 m_W^2\right)B_0(0,m_j^2,m_j^2)+ 2 m_W^4B_0(0,m_W^2,m_W^2)\right.\nonumber\\&+&\left.4 m_j^2 m_W^2-3 m_j^4-m_W^4 \right].\label{Petcov-penguin-Passarino}
\end{eqnarray}

Now, one analytical expression for the $F_a^0$ function can be obtained in a straightforward way either integrating over the Feynman parameters in eq.~(\ref{Petcov-penguin-Feynman}) or using the definition of the $B_0(0,m^2,m^2)$ scalar function in eq.~(\ref{Petcov-penguin-Passarino}). In such a way that after making an expansion around $m_j^2=0$ we obtained

\begin{equation}
F^0_a=\frac{1}{2\pi^2}\left[\frac{m_j^2}{m_W^2}\log\left(\frac{m_W^2}{m_j^2} \right)-\frac{m_j^2}{2m_W^2}+\frac{1}{2}\log\left(\frac{m_W^2}{\mu^2} \right)+\frac{1}{4}+\vartheta\left(\frac{m_j^2}{m_W^2}\right)^2 \right].\label{Fa-Petcov}
\end{equation}
From eq.~(\ref{Fa-Petcov}) it turns clear that, in this approximation, the amplitude will be proportional to the neutrino mass squared, where the dominant contribution, due to the big gap between the neutrino and $W$ boson mass scales, comes from the first term as it involves a relative factor $\log\left(\frac{m_W^2}{m_j^2} \right)$ compared to the second one \footnote{A similar relative suppression operates for the diagrams in fig. \ref{diagrams} (a), (b) and (c) with respect to the diagrams in fig. \ref{diagrams} (d) and (e).}, whereas the independent terms on neutrino mass will vanish by the GIM-like mechanism.

Therefore, the structure of the matrix element for the contribution of the diagram (d) in fig. \ref{diagrams} in the approximation where masses and momenta of the external particles are neglected is given by

\begin{eqnarray}
\mathcal{M}_{d}&=&-i\frac{G_F^2 m_W^2\beta_{F^0_a}}{4}\,\bar{u}(p)\gamma_{\lambda}(1- \gamma_5) u(P) \times \bar{u}(p_{1})\gamma^{\lambda}(1- \gamma_5) v(p_2)\nonumber\\ & +&i G_F^2 m_W^2 s_W^2\beta_{F^0_a}\,\bar{u}(p)\gamma_{\lambda}(1- \gamma_5) u(P) \times \bar{u}(p_{1})\gamma^{\lambda} v(p_2)\,,\label{Md}
\end{eqnarray}
where we have defined


\begin{equation}
\beta_{F^0_a}=\sum_j U_{Lj}U^*_{\ell j}F^0_a(m_j^2).
\end{equation}

We verified that eq.~(\ref{Md}) reproduces the result reported in ref.~\cite{Petcov:1976ff} considering only the first term in eq.~(\ref{Fa-Petcov}) and the simple case of two families.\\

Returning to the general case (non-zero masses and momentum of the external particles), we also have obtained the $F_k$ functions using both Feynman parametrization (we will denote the corresponding expressions by $F_{F_k}$) and the Passarino-Veltman (PaVe) technique (denoted by $F_{PV_k}$) \cite{tHooft:1978jhc, Passarino:1978jh}, employing FeynCalc \cite{Mertig:1990an}. 
In particular, we agree with the expressions previously reported in ref.~\cite{Pham:1998fq} in terms of the Feynman parameters \footnote{We have found some irrelevant differences in the numerators of the $f_d$ and $f_f$ functions, as can be seen comparing eqs. (\ref{IntFeyn1}), (\ref{IntFeyn2}) and  (\ref{IntFeyn3}) with the corresponding expressions in ref.~\cite{Pham:1998fq}.}, namely the $F_{F_k}$ functions can be written as

\begin{equation}
F_{F_k}(q^2, m_j^2)=\frac{1}{2\pi^2}\int_0^1 dx\int_0^{1-x}f_k(q^2, m_j^2)dy,\label{Fey-exp}
\end{equation}
where

\begin{eqnarray}
f_{a}&=& 2 + \log \left(D_j(q^2)/\mu^2\right)+\frac{(q^2-m^2)x(y-1)+M^2x(x+y)+q^2y(y-1)}{D_j},\label{IntFeyn1} \\
f_b&=& \frac{m M x}{D_j},\quad f_c= -\frac{M x (x+y)}{D_j},\quad f_d=-\frac{ m  x (1-y)}{D_j}, \label{IntFeyn2}\\ 
f_e&=&\frac{M x(2-3y-x)-2 M y (y-1)}{D_j}, \quad f_f=\frac{x m (y-1)+2 m y (y-1)}{D_j}, \label{IntFeyn3}
\end{eqnarray} and $D_j$ is defined as

\begin{equation}
D_j(q^2, m_j^2)=-(x-1) m_j^2-m^2xy+x m_W^2+M^2 x (x+y-1)-q^2 y(1-x-y).
\end{equation}

We have omitted in $f_a$ the term associated with the UV divergence since it is independent of $m_j$ and vanishes owing to the GIM-like mechanism. 

On the other hand, the $F_{k}$ functions in terms of the PaVe scalar functions are given as follows

\begin{equation}
F_{PV_k}(q^2, m_j^2)=\frac{1}{2\pi^2}\frac{N_{F_k}}{D_{F_k}},\label{PaVe-exp}
\end{equation}
with
\begin{eqnarray}
D_{F_a}=2 D_{F_b}&=&-2\lambda(m^2,M^2,q^2),\\
D_{F_c}=D_{F_e}&=& \frac{M}{2}D_{F_a}^2 \quad D_{F_d}\,=\,D_{F_f}\,=\,\frac{m}{2}D_{F_a}^2\, ,
\end{eqnarray} 

\begin{eqnarray}
N_{F_k}&=&\xi_{k_1}B_0(m^2,m_j^2,m_W^2)+\xi_{k_2}B_0(M^2,m_j^2,m_W^2)+\xi_{k_3}B_0(q^2,m_j^2,m_j^2)+\xi_{k_4}B_0(0,m_j^2,m_W^2)\nonumber\\& &+\xi_{k_5}C_0(m^2,M^2,q^2,m_j^2,m_W^2,m_j^2)+\xi_{k_0},\label{NumPasarino}
\end{eqnarray}
where $\lambda$ is the Kallen function $\lambda(x,y,z)=x^2+y^2+z^2-2(xy+xz+yz)$, and the $\xi_{k}$ factors can be found in the appendix \ref{Apen}.\footnote{The cancellation of the UV divergences for the $F_m$ functions in terms of the PaVe functions occurs again by the GIM mechanism. This can be verified easily owing to the fact that the sums over the coefficients of the different scalar $B_0$ functions, which contain an isolated divergent term, are independent of $m_j$. That is $\sum_{i=1}^4 \frac{\xi_{a_i}}{D_{PV_a}}=-\frac{1}{2}$, and $\sum_{i=1}^4 \frac{\xi_{l_i}}{D_{PV_l}}=0$ for ($l=b,c,d,e,f$).}



\begin{figure}[!h]
		\centering
		\includegraphics[scale=0.7]{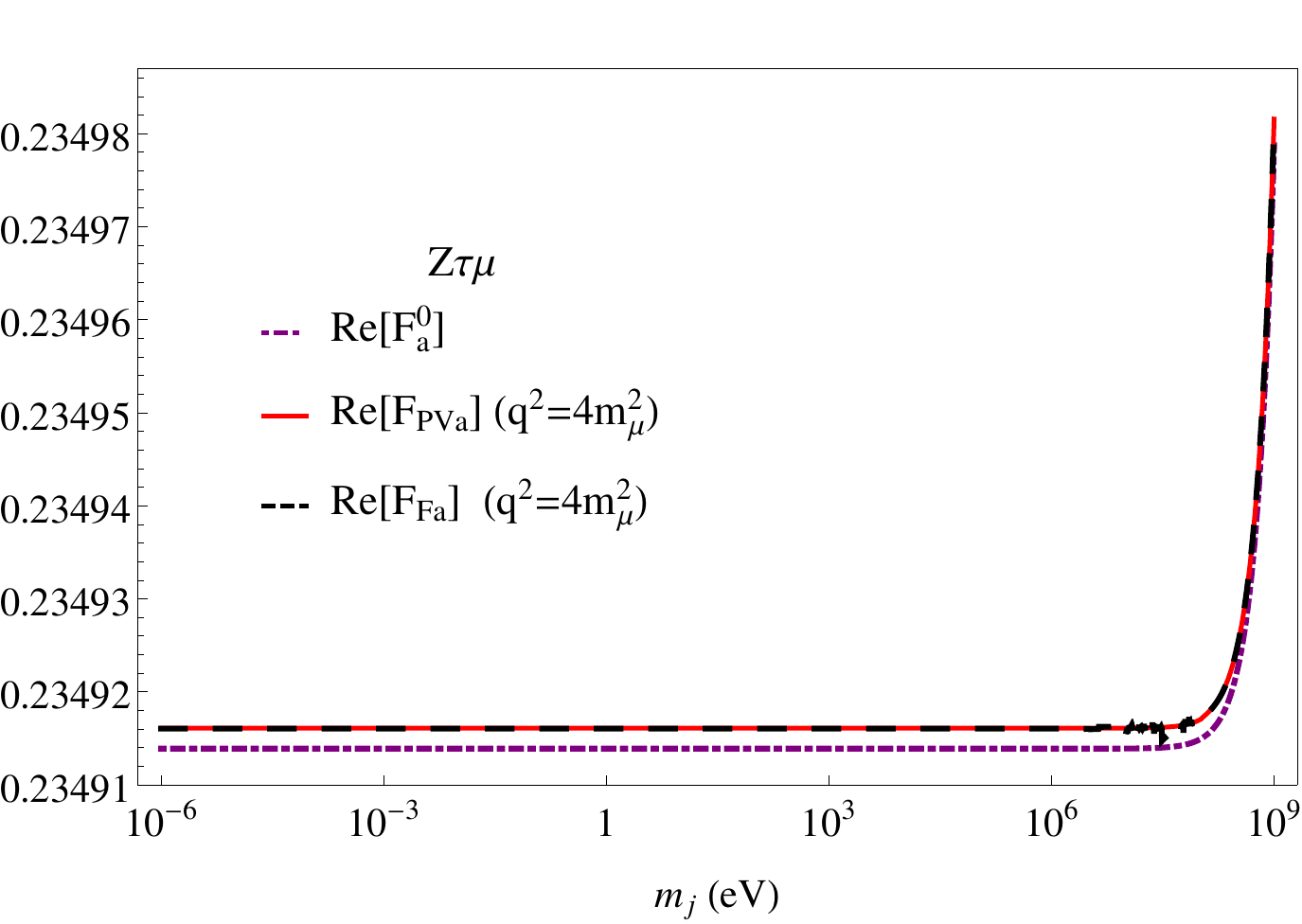}
\caption{Numerical evaluation of the $F_a$ function for the effective $Z\tau\mu$ vertex as a function of the neutrino mass, taking the minimal value of $q^2 = 4m_\mu^2$ for the particular $\tau^- \to\mu^-\mu^-\mu^+$ channel. Black dashed line stands for the numerical evaluation of the $F_a$ function in terms of the Feynman parameters depicted by $F_{F_a}$ (eq.~(\ref{Fey-exp})), whereas the red line corresponds to the evaluation in terms of the PaVe functions represented by $F_{PV_a}$ (eq.~(\ref{PaVe-exp})). We have found some numerical instabilities for the evaluation of the $F_{F_a}$ function in the region $0.01$ GeV$<m_j< 0.1$ GeV. On the other hand, a better precision is achieved in the evaluation of the $F_{PV_a}$ function with the help of the Looptools package. In order to perform a comparison with the approximation done in ref. \cite{Petcov:1976ff}, we also show the complete $F_a^0$ given by the eqs.~(\ref{Petcov-penguin-Feynman}) or (\ref{Petcov-penguin-Passarino}) (purple dotdashed line).}
		\label{Compa}
\end{figure}

  
Unlike the approximation made in ref. \cite{Petcov:1976ff}, the presence of masses and momenta of the external particles in the computation hinders the way for the derivation of analytical expressions for the integrals in eqs.  (\ref{Fey-exp}) or (\ref{PaVe-exp}) ~\footnote{The analytical expressions of the first integrals over the $y$ parameter in eqs. (\ref{IntFeyn1}), (\ref{IntFeyn2}) and (\ref{IntFeyn3}) can be derived from the formulas reported in appendix \ref{Int-y}.}. Nevertheless, we have done a numerical cross-check between both expressions, where we have employed the Looptools package \cite{vanOldenborgh:1989wn, Hahn:1998yk} for the evaluation of the PaVe functions and a numerical Mathematica \cite{Mathematica} routine for the evaluation of the parametric integrals (see fig. \ref{Compa}). We have found an excellent agreement between these two expressions for values of $q^2<4m_j^2$,  which are, however, out side of the physical domain for the considered decays, since $q^2_{\rm{min}}=4m_{\ell'}^2\gg m_j^2$. In this way, owing to the simpler integrals, we verified that a better precision is found in terms of PaVe functions than using Feynman parameters, this feature is illustrated, as an example, for the particular case of the $Z\tau\mu$ transition in fig. \ref{Compa} for the (dominant, as we will show) $F_a$ factor.

At this point, we want to stress that we disagree with the approximation done in ref. \cite{Pham:1998fq} in order to estimate the relevant dependence on the neutrino mass of the $F_k$ functions. We highlight that we are studying a process where the momentum transfer $q^2$ must be non-vanishing and in principle is much larger than the neutrino squared mass, $m_j^2$, which comes from the loop computation. Therefore, using an expansion around $q^2=0$ in order to simplify the integration over the Feynman parameters keeping the terms proportional to $m_j^2$ in the denominators of equations (\ref{IntFeyn1}), (\ref{IntFeyn2}) and (\ref{IntFeyn3}), as it is done in ref.~\cite{Pham:1998fq}, modifies substantially the behavior of the original functions in the interesting physical region for the neutrino masses and, as a consequence, it gives rise to an incorrect infrared logarithmically divergent behavior of the $F_k$ functions when $m_j$ goes to zero, without any possible cure. In particular, the dependence on the momentum transfer, $q^2$, plays a crucial role in the behavior of the $F_k$ functions. In this respect, we point out the presence of a small imaginary part in the $F_a$ function, which emerges for the physical values $4m_j^2<q^2$. 


As we mentioned before, the $q^2$ minimum in the $L^{-}\to \ell^- \ell^{\prime -} \ell^{\prime +}$ decay is given by $4m_{\ell'}^2$, which is much larger than neutrinos masses. This, together with the difficulties in obtaining analytical expressions directly for the $F_k$ functions  suggests employing some numerical approximation to deal with the problem. Because of this, we approximate the $F_k$ functions in the physical region for the neutrinos masses by fitting the curves for the real and imaginary parts of the $F_k$ functions evaluated in terms of the PaVe function \footnote{Our fits for the $F_k$ functions are taken with the precision of the Looptools package considering a neutrino mass varying from $10^{-15}$ GeV to the benchmark point $m_\mu$ ($m_e$), for a fixed value of $q^2 = 4m_\mu^2$ ($q^2 = 4m_e^2$) for the $Z\tau\mu$ ($Z\tau e$ and $Z\mu e$) vertices.}.  We have found a reasonably good fit of the form 

\begin{eqnarray}
F_k&=&\frac{1}{2\pi^2 u}\left(Q_{k}+\frac{m_j^2}{ m_W^2}R_{k}\right),\nonumber\\ 
\label{fit}
\end{eqnarray}
where $u=1$ for $k=a,b$ and $u=M$ for $k=c,d,e,f$  and the respective values for the $Q_k=Q_{R_k}+i Q_{R_I}$ and $R_k=R_{R_k}+i R_{R_I} $ factors  of all considered channels are given in appendix \ref{AdditionalFits}.

From eq. (\ref{fit}), it turns clear that the $Q_k$ factors will not contribute owing to the GIM-like mechanism, whereas the relevant contributions is given by the $R_k$ factors.  
Then, according to our numerical results, we find that the relevant factors of the $F_b$, $F_c$ and $F_d$ functions are suppressed with respect to the $F_a$ factor. On the other hand, despite the respective factors of $F_e$ and $F_f$ functions are larger than those of the $F_a$ function, when the momentum transfer becomes smaller and smaller their helicity suppression makes them negligible. Therefore,  we will concentrate on the contribution of the $F_a$ function. Furthermore, in order to justify our results, we have made an expansion for the PaVe functions involved in eq. (\ref{NumPasarino}), following the same strategy that Cheng and Li for the $\mu\to e\gamma$ decay \cite{LibroCheng}, that is expanding the loop integrals around $m_j^2=0$ (more details of our expansions are given in appendix \ref{Expansion}), and with the help of
Package-X program \cite{Patel:2015tea}, we have been able to rewrite the $F_{PV_a}$ contribution as follows,


\begin{equation}
F_{PV_a}(q^2, m_j^2)=\frac{1}{2\pi^2}\left[Q_a+\frac{m_j^2}{m_W^2}R_a+\vartheta\left(\frac{m_j^4}{m_W^4}\right)\right],\label{Fa-expanded}
\end{equation}

where

\begin{eqnarray}
Q_a&=&-\lambda(m^2,M^2,q^2)^{-1} \left[ f_{Q_{a_1}}C_0(m^2, M^2, q^2, 0,m_W^2,0)+f_{Q_{a_2}}\log \left(\frac{m_W^2}{m_W^2-m^2} \right)\right.\nonumber\\&+&\left. f_{Q_{a_3}}\log \left(\frac{m_W^2}{m_W^2-M^2} \right)+f_{Q_{a_4}}\log \left(\frac{m_W^2}{q^2} \right)+f_{Q_{a_5}}\right]-\frac{1}{2} \Delta,\label{Qa}
\end{eqnarray}

\begin{eqnarray}
R_a&=&-m_W^2\lambda(m^2,M^2,q^2)^{-1} \left[ f_{R_{a_1}}C_0(m^2, M^2, q^2, 0,m_W^2,0)+f_{R_{a_2}}\log \left(\frac{m_W^2}{m_W^2-m^2} \right)\right.\nonumber\\&+&\left. f_{R_{a_3}}\log \left(\frac{m_W^2}{m_W^2-M^2} \right)+f_{R_{a_4}}\log \left(\frac{m_W^2}{q^2} \right)+f_{R_{a_5}}\right],\label{R_a}
\end{eqnarray}
where $\Delta=\frac{1}{\textstyle{\epsilon}}-\gamma_E+\log(4\pi)$, and the $f_Q$ and $f_R$ factors can be found in the appendix \ref{Expansion}. We verified that our numerical fits for the $Z\tau\mu$ and $Z\tau e$ vertex are in a very good agreement with eq. (\ref{R_a}), whereas a deviation is found for the $Z\mu e$ vertex, as can be seen in Table \ref{Compariso Ra}, we consider the results obtained from eq. (\ref{R_a}) for the effective vertices as our reference ones.

In this way, we can approximate the amplitude for diagram (d) according to eq.~(\ref{Md}) replacing $F_a^0$ by

\begin{equation}
F_a\approx \frac{1}{2\pi^2}\frac{m_j^2}{m_W^2}R_a,
\end{equation}


Now, in order to evaluate the respective branching fractions for the $L^{-}\to \ell^- \ell^{\prime-} \ell^{\prime +}$ decays we considered the state of the art best fit values of the three neutrino oscillation parameters \cite{Patrignani:2016xqp, Fits}. Without lose of generality, we assume the $CP$-conserving scenario~\footnote{In general, the leptonic mixing matrix can involve three $CP$-violating phases, one Dirac phase $\delta$, and two additional physical phases in case neutrinos are Majorana particles. 
Lepton number conserving observables (as those considered here) are not sensitive to the latter, so that for $L^{-}\to \ell^- \ell^{\prime -} \ell^{\prime +}$ decays they can only depend on the phase $\delta$. Once the unitarity condition has been used to write the lightest neutrino mass eigenstate contribution in terms of the other two, it can be seen that the term with the largest logarithm (Log$(m_3^2/m_1^2)$ in the normal hierarchy) has a PMNS pre-factor (we are using the PDG parametrization) which does not depend on $\delta$, which justifies our approach.}, and we use the following values reported 
for the mixing angles $\sin^2\theta_{12}=0.307(13)$, $\sin^2\theta_{23}=0.51(4)$, and $\sin^2\theta_{13}=0.0210(11)$, whereas the neutrino mass squared differences are taken as $\Delta m^2_{32}=2.45(5)\times 10^{-3}$eV$^2$ and $\Delta m^2_{21}=7.53(18)\times 10^{-5}$eV$^2$ \footnote{These numbers correspond to the normal hierarchy ($m_1<m_2<m_3$); different (though very similar) values are reported for the inverted hierarchy ($m_3<m_1<m_2$). Changing hierarchy is immaterial for our numerical evaluations. We have verified that results are not sensitive to the lightest neutrino mass 
value, but only to the mass squared differences.}. The kinematics for the $L^{-}\to \ell^- \ell^{\prime -} \ell^{\prime +}$ decays  can be found in Appendix \ref{Kinematics}.

Neglecting for the moment the box contributions, we get the branching fractions reported in table \ref{tab:BRs}. 


\begin{table}[h!]
\begin{center}
\begin{tabular}{|c|c|c|}
\hline
Decay channel & Our result & Ref.~\cite{Petcov:1976ff} \tabularnewline
\hline 
$\mu^-\to e^- e^+ e^-$ & $9.5\cdot10^{-55}$ & $1.0\cdot10^{-53}$\tabularnewline
\hline 
$\tau^-\to e^- e^+ e^-$ & $5.0\cdot10^{-56}$& $1.8\cdot10^{-54}$\tabularnewline
\hline
$\tau^-\to \mu^- \mu^+ \mu^-$ & $1.0\cdot10^{-54}$  & $3.7\cdot10^{-53}$\tabularnewline
\hline 
$\tau^-\to e^- \mu^+ \mu^-$ & $2.9\cdot10^{-56}$ & $1.0\cdot10^{-54}$\tabularnewline
\hline 
$\tau^-\to \mu^- e^+ e^-$ & $7.3\cdot10^{-55}$ & $2.5\cdot10^{-53}$\tabularnewline

\hline
\end{tabular}
\par\end{center}
\caption{Branching ratios for the $L^{-}\to \ell^- \ell^{\prime -} \ell^{\prime +}$ decays (neglecting the box and the subdominant penguin contributions with only one neutrino propagator), which are obtained using the current knowledge of the PMNS matrix. The last column values correspond to the approximation where external masses and momenta are neglected \cite{Petcov:1976ff}. Our results are smaller than those by around one (two) orders of magnitude for the $\mu$ ($\tau$) decays.}\label{tab:BRs}
\end{table}

\section{Contributions of the box diagrams}

Now, in order to make a complete comparison with the approximation done in ref.~\cite{Petcov:1976ff} we have also obtained the amplitude for the box diagram (e) in fig. \ref{diagrams}. Note that unlike the penguin diagram (d), which involves two neutrino propagators of the same flavor, the box diagram (e) can involve two neutrino propagators with different flavors. Thus, in full generality, the amplitude can be written as follows

  \begin{equation}
  \mathcal{M}_{e}=\left(\frac{-ig}{2\sqrt{2}}\right)^4\sum_{i,j} U_{Lj}U^*_{lj} U_{\ell^\prime i}U^*_{\ell^\prime i}T_{\sigma\sigma'}I^{\sigma\sigma'}, 
   \end{equation}  
where we defined 

\begin{eqnarray}
T_{\sigma\sigma'}&=&4\,\bar{u}(p)\gamma_\mu\gamma_\sigma\gamma_\nu(1-\gamma_5)u(P)\times\bar{u}(p_1)\gamma^\nu\gamma_{\sigma'}\gamma^\mu(1-\gamma_5)v(p_2)\label{Tensor}
\end{eqnarray}  
and the relevant loop integral is given by (see fig. \ref{diagrams} (e))

\begin{equation}
I^{\sigma\sigma'}=\int\frac{d^4k}{(2\pi)^4}\frac{(P+k)^\sigma(k+p_1)^{\sigma'}}{(k^2-m_W^2)[(p_1+p_2+k)^2-m_W^2][(P+k)^2-m_j^2][(k+p_1)^2-m_i^2]}.\label{Int-Box}
\end{equation}
Since we have written the equation (\ref{Int-Box}) in terms of $P$, $p_1$ and $p_2$ momenta the integral must take the form 

\begin{eqnarray}
I^{\sigma\sigma'}&=&i\left(g^{\sigma\sigma'}H_a+P^{\sigma}P^{\sigma'}H_b+P^{\sigma}p_1^{\sigma'}H_c+P^{\sigma}p_2^{\sigma'}H_d+p_1^{\sigma}P^{\sigma'}H_e\right. \nonumber\\&+&\left. p_1^{\sigma}p_1^{\sigma'}H_f+p_1^{\sigma}p_2^{\sigma'}H_g+p_2^{\sigma}P^{\sigma'}H_h+p_2^{\sigma}p_1^{\sigma'}H_i+p_2^{\sigma}p_2^{\sigma'}H_j \right)\label{Itotal}. 
\end{eqnarray} The $H_k$ factors depend upon the kinematical variables $s_{12}=(p_1+p_2)^2=q^2$ and $s_{13}=(p_1+p)^2$, in addition of $m_i$ and $m_j$.

Anew, in the approximation where momenta of the external particles are neglected in eq. (\ref{Int-Box}), the only contribution is given by the $H^0_a$ function, which  will not depend either on $s_{12}$ or $s_{13}$. In such case, we obtained the following simplified expression

\begin{equation}
H_{a}^0(m_j^2, m_i^2)=\frac{1}{16\pi^2}\int_0^1 dx\int_0^{1-x}dy\int_0^{1-x-y}\frac{-1}{2M_{F_0}^2}dz, \label{Box-Petcov-Feynman}
\end{equation}where

\begin{eqnarray}
M_{F_{0}}^2&=&m_W^2(x+y)-m_j^2(x+y-1)+(m_i^2-m_j^2)z.
\end{eqnarray}

Whereas, in terms of PaVe functions, $H^0_a$ reads

\begin{eqnarray}
H^0_{a}(m_j^2, m_i^2)&=&\frac{1}{16\pi^2}\left(\frac{m_j^4}{4 \left(m_j^2-m_i^2\right) \left(m_j^2-m_W^2\right){}^2}B_0(0,m_j^2,m_j^2)+\frac{m_i^4}{4 \left(m_i^2-m_j^2\right) \left(m_i^2-m_W^2\right){}^2}B_0(0,m_i^2,m_i^2)\right.\nonumber\\&+&\left.\frac{2 m_i^2 m_j^2 m_W^2-m_W^4 \left(m_i^2+m_j^2\right)}{4 \left(m_i^2-m_W^2\right){}^2 \left(m_j^2-m_W^2\right)^2}B_0(0,m_W^2,m_W^2)+\frac{m_W^2}{4 \left(m_i^2-m_W^2\right) \left(m_W^2-m_j^2\right)}\right)\,. \label{Box-Petcov-Passarino}
\end{eqnarray}

In the same way that $F_a^0$ form factor, an analytical expression for $H_a^0$ can be obtained easily from either eq.~(\ref{Box-Petcov-Feynman}) or eq.~(\ref{Box-Petcov-Passarino}). This time, making a double Taylor expansion, first around $m_i^2=0$ and then around $m_j^2=0$, we obtained that

\begin{eqnarray}
H^0_a(m_j^2, m_i^2)&=&\frac{1}{64\pi^2 m_W^4}\left[\left(m_i^2+m_j^2 \right)\left(\log\left(\frac{m_W^2}{m_j^2}\right)-1\right)\right.\nonumber\\&+&\left.\frac{m_i^2m_j^2}{m_W^2}\left(2\log\left(\frac{m_W^2}{m_j^2}\right)-1\right) -m_W^2+\vartheta\left(\frac{m_i^4}{m_W^2}\right)+\vartheta\left(\frac{m_j^4}{m_W^2}\right)\right].\label{H0aPetcov}
\end{eqnarray}

Using that $T_{\sigma\sigma'}g^{\sigma\sigma'}=16\bar{u}(p)\gamma_{\lambda}(1- \gamma_5) u(P) \times \bar{u}(p_{1})\gamma^{\lambda}(1- \gamma_5) v(p_2)$, the amplitude -in this approximation- is given by

\begin{eqnarray}
\mathcal{M}_{e}&=&i8G_F^2 m_W^4\beta_{H^0_a}\,\bar{u}(p)\gamma_{\lambda}(1- \gamma_5) u(P) \times \bar{u}(p_{1})\gamma^{\lambda}(1- \gamma_5) v(p_2),\label{Me}
\end{eqnarray}
with
\begin{equation}
\beta_{H^0_a}=\sum_{j,i} U_{Lj}U^*_{\ell j}  U_{\ell' i}U^*_{\ell' i} H^0_a(m_i^2, m_j^2).
\end{equation}

Again, we verified that taking into account the first term in eq.~(\ref{H0aPetcov}) and considering only two families, eq.~(\ref{Me}) reproduces the expression reported in ref.~\cite{Petcov:1976ff} for the amplitude of the box diagram \ref{diagrams} (e) in the $\mu\to 3e$ decay.

In the general case, we also obtained the $H_k$ ($k=a,b,...,j$) functions in terms of both Feynman parameters integrals, $H_{F_k}$, and PaVe functions, $H_{PV_k}$. This time, the $H_k$ functions will depend on the squared masses of two different neutrinos, $m_j^2$ and $m_i^2$, and on two independent phase space variables $s_{12}$ and $s_{13}$. Using Feynman parametrization these functions read

\begin{equation}
H_{F_k}(s_{12},s_{13},m_i^2, m_j^2)=\frac{1}{16\pi^2}\int_0^1 dx\int_0^{1-x}dy\int_0^{1-x-y}h_k(s_{12},s_{13},m_i^2, m_j^2)dz \,,\label{H_k Feynman}
\end{equation}

where

\begin{equation}
h_a=-\frac{1}{2M_F^2},\quad h_b=\frac{z(z-1)}{M_F^4}, \quad h_c=-\frac{(z-1)(x+z)}{M_F^4},\quad h_d=\frac{y(z-1)}{M_F^4}\quad h_e=-\frac{z(x+z-1)}{M_F^4}\,,
\end{equation}

\begin{equation}
 h_f=\frac{(x+z-1)(x+z)}{M_F^4}, \quad h_g=-\frac{y(x+z-1)}{M_F^4}, \quad h_h=\frac{yz}{M_F^4}\quad 
h_i=-\frac{y(x+z)}{M_F^4}, \quad h_j=\frac{y^2}{M_F^4}.
\end{equation}

In the previous expressions, the denominator function is given by

\begin{eqnarray}
M_F^2&=&-m_j^2 (x+y-1)+m_{\ell'}^2 (x+y-1) (x+y)+ m_W^2(x+y)-s_{12} x y+z^2 \left(2 m_{\ell'}^2+m^2+M^2-s_{12}-s_{13}\right)\nonumber\\&+&z \left[m_i^2-m_j^2+(x+y) \left(3 m_{\ell'}^2-s_{12}-s_{13}\right)-2 m_{\ell'}^2+m^2 (x-1)+M^2 (y-1)+s_{12}+s_{13}\right].
\end{eqnarray}

Expressions are rather lengthy in terms of the PaVe functions, so that here we only present the expression for the dominant $H_a$ function, which can be written as

\begin{equation}
H_{PV_a}(s_{12}, s_{13}, m_j^2, m_i^2)=\frac{1}{16\pi^2}\frac{N_{H_a}}{D_{H_a}},\label{PaVe-Ha-exp}
\end{equation}
with
\begin{eqnarray}
D_{H_a}&=&4 \big(m^4 m_{\ell'}^2-m^2 \big[M^2 \big(2 m_{\ell'}^2-s_{12}\big)+s_{12} \big(m_{\ell'}^2+s_{13}\big)\big]+M^4 m_{\ell'}^2-M^2 s_{12} \big(m_{\ell'}^2+s_{13}\big)\nonumber\\&+&s_{12} \big(-2 s_{13} m_{\ell'}^2+m_{\ell'}^4+s_{13} \big(s_{12}+s_{13}\big)\big)\big),
\end{eqnarray} 
and
\begin{eqnarray}
N_{H_a}&=&\chi_{k_1}C_0(m^2,M^2,s_{12},m_W^2,m_i^2,m_W^2)+\chi_{k_2}C_0(m_{\ell'}^2,m_{\ell'}^2,s_{12},m_W^2,m_j^2,m_W^2)\nonumber\\&+&\chi_{k_3}C_0(M^2, m_{\ell'}^2, m^2 + M^2 + 2m_{\ell'}^2 - s_{12} - s_{13}, m_i^2, m_W^2, m_j^2)\nonumber\\&+&\chi_{k_4}C_0(m^2, m_{\ell'}^2, m^2 + M^2 + 2m_{\ell'}^2 - s_{12} - s_{13}, m_i^2, m_W^2, m_j^2)\nonumber\\&+ &\chi_{k_5}D_0(m^2, M^2, m_{\ell'}^2, m_{\ell'}^2, s_{12}, m^2 + M^2 + 2m_{\ell'}^2 - s_{12} - s_{13}, m_W^2, m_i^2,  m_W^2, m_j^2).\label{NumPasarino1}
\end{eqnarray} where $\chi_{k}$ factors are reported in Appendix \ref{Apen}.

\begin{figure}[!h]
		\centering
		\includegraphics[scale=0.7]{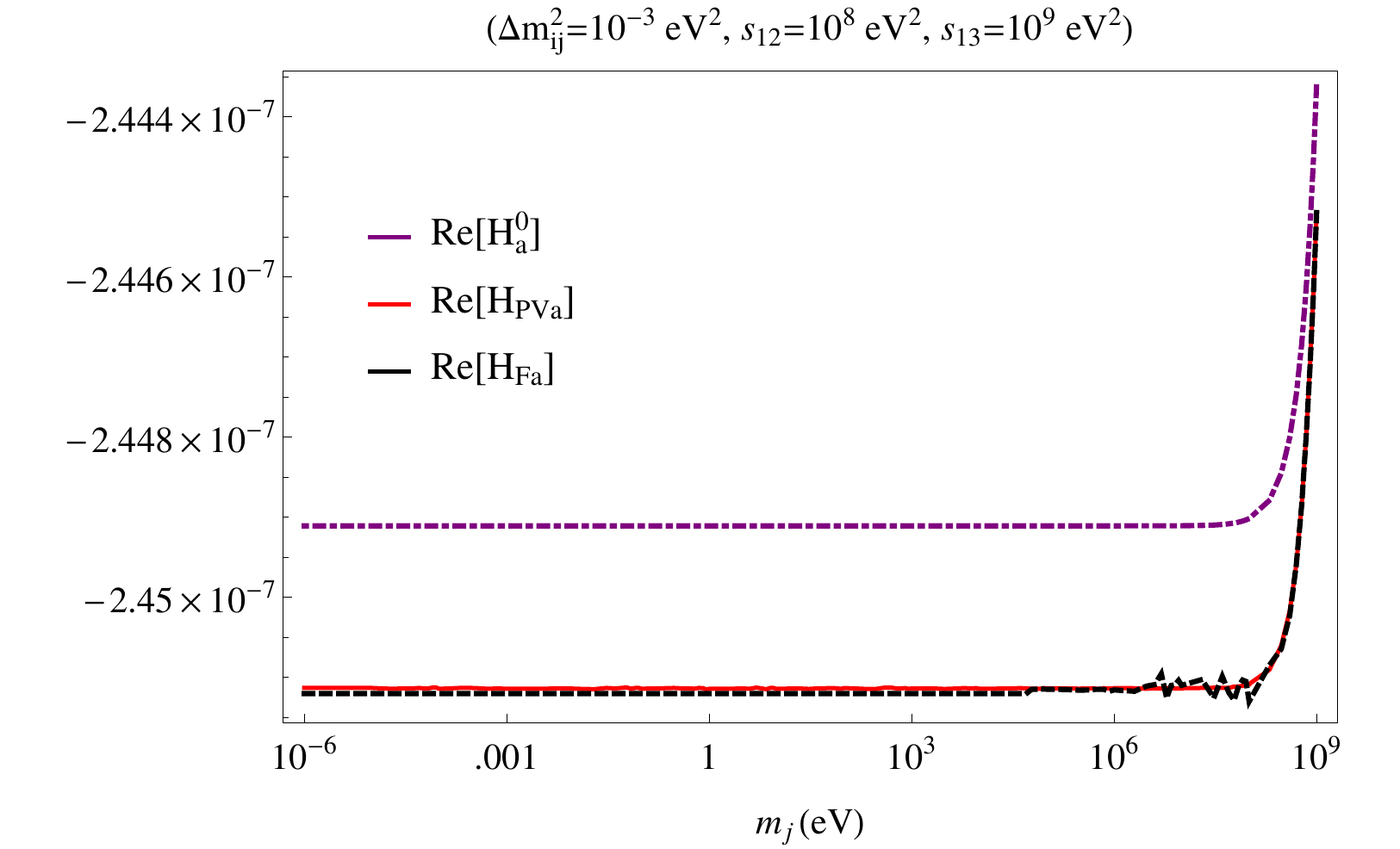}
		\caption{Numerical evaluation of the $H_a$ function versus the neutrino mass. We are considering that $\Delta m_{ij}^2=10^{-3}$ eV$^2$ and the values of $s_{12}=10^{8}$ eV$^2$ and $s_{13}=10^{9}$ eV$^2$ associated with a representative point in the physical phase space for the particular  $\tau^-\to\mu^-\mu^-\mu^+$ channel. In analogous way to the fig. \ref{Compa}, black dashed line stands for the numerical evaluation of the $H_a$ function in terms of the Feynman parameters depicted by $H_{F_a}$ (eq.~(\ref{H_k Feynman})), whereas the red line corresponds to the evaluation in terms of the PaVe functions represented by $H_{PV_a}$ (eq.~(\ref{PaVe-Ha-exp})). Numerical instabilities for the evaluation of the $H_{F_a}$ function around $0.001$ GeV$<m_j<1$ GeV are found. A better precision is achieved for the evaluation of the $F_{PV_a}$ function with the help of the Looptools package. In order to perform a comparison with the approximation done in ref.~\cite{Petcov:1976ff}, we also show the complete $H_a^0$ given by the eqs.~(\ref{Box-Petcov-Feynman}) or (\ref{Box-Petcov-Passarino}) (purple dotdashed line).}
		\label{CompaHa}
\end{figure}

As far as the general case is concerned, we can see that although there are additional contributions associated with the $H_k$ functions, with $k=b$, $c$, $d$, $\ldots$  $j$; they are expected to be suppressed, as they correspond to higher-dimensional operators, with respect  to the $H_a$ function associated with a $(V-A)\times(V-A)$ operator. Therefore, we will concentrate on the $H_a$ function in order to estimate the box diagram contribution. We also have done a numerical cross-check between the expressions for the $H_a$ function given in terms of the Feynman parameters eq.~(\ref{H_k Feynman}) and the PaVe functions eq.~(\ref{PaVe-Ha-exp}), as can be seen in fig. \ref{CompaHa}.
In this case, it turns very complicated and far away of the purpose of this work to obtain an analytical expression for the $H_a$ function in eq. (\ref{NumPasarino1}) making an expansion for the respective scalar PaVe functions, owing to the number of propagators involved and the dependence on two different neutrino masses. However, we can expect a good approximation through our numerical results, such as occurs with the penguin contribution.

Thus, we estimate the relevant dependence on the neutrino mass for the $H_a$ function fitting the curve for the real and imaginary parts of the $H_a$ function evaluated in terms of the PaVe functions considering fixed values for the $m_i$, $s_{12}$, and $s_{13}$ parameters \footnote{Our fits for the $H_a$ function are taken considering an interval for the neutrino mass varying from $10^{-15}$ GeV to $10$ GeV.}. We obtained a good fit of the form 

\begin{eqnarray}
H_a&=&\frac{1}{16\pi^2}\left(Q_{H_a}+\frac{m_j^2}{ m_W^4}R_{H_a}\right), \label{fitHa}
\end{eqnarray}where $R_{H_a}\approx1.5+i0.007 $,  for all different $\tau\to\ell^-\ell^{\prime -}\ell^{\prime +}$ channels, whereas $R_{H_a} \approx 1.5$, for the $\mu^-\to e^-e^-e^+$ channel. These numbers were obtained considering that $\Delta m^2_{ij}=10^{-3}$ eV$^2$,  and representative values for $s_{12}$ and $s_{13}$ within the corresponding phase space.

Now we can evaluate the branching ratios for the  $L^{-}\to \ell^- \ell^{\prime -} \ell^{\prime +}$ decays using the previous results. We will first make a partial evaluation neglecting the penguin contributions (only box diagrams are considered), which yields the values in table \ref{BRbox}.

\begin{table}[h!]
\centering
\begin{tabular}{|l|l|l|}
\hline
Decay channel          & Our Result   & Ref. \cite{Petcov:1976ff}  \\ \hline
$\mu^-\to e^-e^+e^-$    & $2.1 \cdot 10^{-56} $    & $2.6 \cdot 10^{-53}$  \\ \hline
$\tau^-\to e^-e^+e^-$   & $3.6 \cdot 10^{-57}$   &  $4.5 \cdot 10^{-54}$  \\ \hline
$\tau^-\to\mu^-\mu^+\mu^-$& $7.6 \cdot 10^{-56}$   &  $9.7 \cdot 10^{-53}$  \\ \hline
$\tau^-\to e^-\mu^+\mu^-$& $1.7 \cdot 10^{-57} $   &  $2.2 \cdot 10^{-54}$  \\ \hline
$\tau^-\to \mu^-e^+e^-$& $4.0 \cdot 10^{-56}$   &  $5.0 \cdot 10^{-53}$  \\ \hline
\end{tabular}
\caption{Branching ratios for the $L^{-}\to\ell^- \ell^{\prime -} \ell^{\prime +} $ decays (neglecting the penguin contributions), which are obtained using the current knowledge of the PMNS matrix. Our results are obtained taking into account only the contribution from the dominant $H_a$ function. The last column values correspond to the approximation where external masses and momenta are neglected \cite{Petcov:1976ff}. Our results are smaller than those by three orders of magnitude, approximately.}\label{BRbox}
\end{table}


Our final results, where the dominant penguin and box contributions are considered, are collected in table \ref{BRIn}, where they are compared to those obtained using Petcov's results  \cite{Petcov:1976ff} with updated input.  Our predictions are even smaller than Petcov's updated results, as a consequence of keeping external masses and momenta in our computations.

\begin{table}[h!]
\centering
\begin{tabular}{|l|l|l|}
\hline
Decay channel           & Our Result  & Ref. \cite{Petcov:1976ff} \\ \hline
$\mu^-\to e^-e^+e^-$     &  $7.4 \cdot 10^{-55}$    & $8.5 \cdot 10^{-54}$ \\ \hline
$\tau^-\to e^-e^+e^-$    & $3.2 \cdot 10^{-56}$  &  $1.4 \cdot 10^{-54}$ \\ \hline
$\tau^-\to\mu^-\mu^+\mu^-$& $6.4 \cdot 10^{-55}$   &  $3.2 \cdot 10^{-53}$  \\ \hline
$\tau^-\to e^-\mu^+\mu^-$& $2.1 \cdot 10^{-56}$  &  $9.4 \cdot 10^{-55}$  \\ \hline
$\tau^-\to \mu^-e^+e^-$  & $5.2 \cdot 10^{-55}$   &  $2.1 \cdot 10^{-53}$ \\ \hline
\end{tabular}
\caption{Branching ratios including all contributions (interferences are not neglected), which are obtained using the current knowledge of the PMNS matrix. Our results are obtained taking into account only the contribution from the dominant $H_a$ function. The last column values correspond to the approximation where external masses and momenta are neglected \cite{Petcov:1976ff}. Our results are smaller than those by around one (two) orders of magnitude for the $\mu$ ($\tau$) decays}\label{BRIn}
\end{table}


These extremely suppressed branching ratios for lepton flavor violating $L^{-}\to \ell^- \ell^{\prime -} \ell^{\prime +}$ decays due to massive light neutrinos are found at similar rates in the case of LFV Z \cite{Illana:2000ic} and Higgs boson decays \cite{Arganda:2004bz}.

\section{Conclusions}\label{Concl}

We have revisited the $L^-\to \ell^-\ell^{\prime -}\ell^{\prime +}$ decays in the SM with massive neutrinos.  We obtained expressions in terms of both Feynman parameters and scalar Passarino-Veltman functions for the relevant loop integrals of the (dominant) diagrams that involve two neutrino propagators considering non-vanishing masses and momenta of the external particles. Opposed to the previous calculation reported in ref. \cite{Pham:1998fq}, we found that all the different amplitudes for these processes are strongly suppressed (as they are proportional to the neutrino mass squared). In the particular case of the penguin contribution with two neutrino propagators, we highlight that it is crucial to save the dependence on the momentum transfer in the Feynman integrals in order to evaluate the amplitude in the physical region for the neutrino masses. This fact avoids the incorrect divergent logarithmic behavior in the amplitude claimed in ref. \cite{Pham:1998fq}. As far as the box contribution is concerned, we found that the dominant term comes from $H_a$ function that is associated with a (V-A)$\times$(V-A) operator,  and it is in  good agreement with the approximation done in Ref. \cite{Petcov:1976ff}.\\

Current and forthcoming experiments were approaching the limits predicted by ref. \cite{Pham:1998fq} on the SM prediction for the lepton flavor violating $\tau^-\to \mu^- \ell^+ \ell^-$ ($\ell=\mu, e$) decays due to non-zero neutrino masses. This prediction was at odds with ref. \cite{Petcov:1976ff} corresponding computation for the $\mu^-\to e^- e^+ e^-$ decays predicting an extremely suppressed, unmeasurable branching ratio  (as in $L^-\to \ell^- \gamma$ processes). The most important result of our analysis is the confirmation (in agreement with ref. \cite{Petcov:1976ff}) that any future observation of $L^{-}\to \ell^- \ell^{\prime -} \ell^{\prime +}$ decays would imply the existence of New Physics.

\acknowledgments{The authors are indebted to Swagato Banerjee and Simon Eidelman for pointing us the interest of this calculation. We are thankful to Serguey Petcov for fruitful discussions. Finally, we also acknowledge support from Conacyt through projects FOINS-296-2016 (‘Fronteras de la Ciencia’), and 236394 and 250628 (Ciencia B\'asica).}

\appendix

\section{One-loop PaVe scalar functions}\label{Apen}

In this appendix we collect the $\left\lbrace\xi_{i_j}\right\rbrace_{{i=a,...,f; j=0,..,5}}$ factors entering our results in eq.~(\ref{NumPasarino}):

\begin{eqnarray}
\xi_{a_0}&=& D_{F_a}, \\
\xi_{a_1}&=& -m^2 \left(m_j^2-m_W^2+M^2+q^2\right)+\left(M^2-q^2\right) \left(m_j^2-m_W^2+2 M^2-2 q^2\right)-m^4, \\
\xi_{a_2}&=& -q^2 \left(m_j^2+4 m^2-m_W^2+M^2\right)+\left(m^2-M^2\right) \left(m_j^2+2 m^2-m_W^2+M^2\right)+2 q^4, \\ 
\xi_{a_3}&=& q^2 \left(2 m_j^2+3 m^2-2 m_W^2+3 M^2-3 q^2\right), \\
\xi_{a_4}&=&0, \\
\xi_{a_5}&=&-2 q^2 \left(m^2 \left(2 m_j^2-2 m_W^2+M^2-2 q^2\right)+\left(m_j^2-m_W^2+M^2-q^2\right){}^2+m^4\right).
\end{eqnarray}

\begin{eqnarray}
\xi_{b_0}&=& \xi_{b_4}=0, \\
\xi_{b_1}&=& -m M \left(m^2-M^2+q^2\right), \\
\xi_{b_2}&=& m M \left(m^2-M^2-q^2\right), \\ 
\xi_{b_3}&=&  2 m M q^2,\\
\xi_{b_5}&=&  -m M q^2 \left(2 m_j^2+m^2-2 m_W^2+M^2-q^2\right).
\end{eqnarray}

\begin{eqnarray}
\xi_{c_0}&=&M^2 \big(-m^6+m^4 \big(3 M^2+q^2\big)+m^2 \big(-3 M^4+2 M^2 q^2+q^4\big)+\big(M^2-q^2\big)^3\big), \\
\xi_{c_1}&=& M^2 \big(-m^4 \big(m_j^2-m_W^2+4 M^2+6 q^2\big)+m^2 \big(2 M^2 \big(m_j^2-m_W^2-4 q^2\big)+q^2 \big(-10 m_j^2+10 m_W^2+3 q^2\big)+5   M^4\big)\nonumber\\&&-\big(M^2-q^2\big)^2 \big(m_j^2-m_W^2+2 M^2-2 q^2\big)+m^6\big),  \\
\xi_{c_2}&=& -q^4 \big(m^2 \big(3 m_j^2-3 m_W^2+7 M^2\big)+2 M^2 \big(3 m_j^2-3 m_W^2+2 M^2\big)\big)-\big(m^2-2 M^2\big) \big(m^2-M^2\big)^2
   \big(m_j^2-m_W^2+M^2\big)\nonumber\\&&+q^2 \big(m^4 \big(3 m_j^2-3 m_W^2+5 M^2\big)+2 m^2 M^2 \big(m_j^2-m_W^2+2 M^2\big)\nonumber\\&&-M^4 \big(-3 m_j^2+3
   m_W^2+M^2\big)\big)+q^6 \big(m_j^2-m_W^2+3 M^2\big), \\
 \xi_{c_3}&=&M^2 q^2 \big(m^2 \big(6 m_j^2-6 m_W^2+4 M^2+4 q^2\big)-\big(M^2-q^2\big) \big(6 m_j^2-6 m_W^2+5 M^2-5 q^2\big)+m^4\big) ,\\  
\xi_{c_4}&=& \big(m_j^2-m_W^2\big) \big((m-M)^2-q^2\big) \big(m^2-M^2-q^2\big) \big((m+M)^2-q^2\big), \\
\xi_{c_5}&=& -2 M^2 \big(m^6 m_j^2+m^4 \big(M^2 \big(2 q^2-3 m_j^2\big)+q^2 \big(m_j^2-2 m_W^2+q^2\big)\big)+m^2 \big(M^4 \big(3 m_j^2-q^2\big)+q^2
   \big(q^2 \big(m_j^2-2 m_W^2\big)\nonumber\\&&+3 \big(m_j^2-m_W^2\big){}^2-2 q^4\big)+M^2 q^2 \big(3 q^2-2 m_W^2\big)\big)-\big(M^2-q^2\big)
   \big(M^4 \big(m_j^2+q^2\big)\nonumber\\&&+2 M^2 q^2 \big(m_j^2-2 m_W^2-q^2\big)+q^2 \big(-3 q^2 m_j^2+3 \big(m_j^2-m_W^2\big){}^2+4 q^2 m_W^2+q^4\big)\big)\big).
\end{eqnarray}

\begin{eqnarray}
\xi_{d_0}&=& m^2 \big(m^6-3 m^4 \big(M^2+q^2\big)+m^2 \big(3 M^4+2 M^2 q^2+3 q^4\big)-\big(M^2-q^2\big)^2 \big(M^2+q^2\big)\big), \\
\xi_{d_1}&=& -m^6 \big(-2 m_j^2+2 m_W^2+5 M^2+q^2\big)+m^4 \big(M^2 \big(-5 m_j^2+5 m_W^2+4 q^2\big)+q^2 \big(3 m_j^2-3 m_W^2-4 q^2\big)+4 M^4\big)\nonumber\\&&-m^2
   \big(M^2-q^2\big) \big(-4 M^2 \big(m_j^2-m_W^2+q^2\big)+3 q^2 \big(-2 m_j^2+2 m_W^2+q^2\big)+M^4\big)\nonumber\\&&-\big(M^2-q^2\big)^3
   \big(m_j^2-m_W^2\big)+2 m^8, \\
\xi_{d_2}&=& m^2 \big(q^4 \big(-m_j^2-6 m^2+m_W^2+3 M^2\big)-\big(m^2-M^2\big)^2 \big(m_j^2+2 m^2-m_W^2-M^2\big)\nonumber\\&&+2 q^2 \big(m^2 \big(m_j^2-m_W^2-4
   M^2\big)+M^2 \big(-5 m_j^2+5 m_W^2-3 M^2\big)+3 m^4\big)+2 q^6\big), \\ 
\xi_{d_3}&=& m^2 q^2 \big(2 q^2 \big(3 m_j^2+5 m^2-3 m_W^2+2 M^2\big)-\big(m^2-M^2\big) \big(6 m_j^2+5 m^2-6 m_W^2+M^2\big)-5 q^4\big), \\
\xi_{d_4}&=& \big(m_j^2-m_W^2\big) \big(-\big((m-M)^2-q^2\big)\big) \big((m+M)^2-q^2\big) \big(m^2-M^2+q^2\big), \\
\xi_{d_5}&=&2 m^2 \big(m^6 \big(m_j^2+q^2\big)+m^4 \big(M^2 \big(q^2-3 m_j^2\big)+q^2 \big(m_j^2-4 m_W^2-3 q^2\big)\big)+m^2 \big(M^4 \big(3 m_j^2-2
   q^2\big)\nonumber\\&&+q^2 \big(-5 q^2 m_j^2+3 \big(m_j^2-m_W^2\big){}^2+8 q^2 m_W^2+3 q^4\big)+M^2 q^2 \big(2 m_W^2-3 q^2\big)\big)-M^6 m_j^2-M^4
   q^2 \big(m_j^2-2 m_W^2+q^2\big)\nonumber\\&&+M^2 q^2 \big(-q^2 \big(m_j^2-2 m_W^2\big)-3 \big(m_j^2-m_W^2\big){}^2+2 q^4\big)\nonumber\\&&+q^4 \big(3 m_j^2
   \big(2 m_W^2+q^2\big)-3 m_j^4-\big(m_W^2+q^2\big) \big(3 m_W^2+q^2\big)\big)\big).
\end{eqnarray}

\begin{eqnarray}
\xi_{e_0}&=&-M^2 \big(3 m^6-m^4 \big(5 M^2+7 q^2\big)+m^2 \big(M^4-6 M^2 q^2+5 q^4\big)+\big(M^2-q^2\big)^3\big),  \\
\xi_{e_1}&=& M^2 \big(m^4 \big(-11 m_j^2+11 m_W^2+2 q^2\big)+m^2 \big(2 M^2 \big(5 m_j^2-5 m_W^2-4 q^2\big)+q^2 \big(-2 m_j^2+2 m_W^2+5 q^2\big)+3
   M^4\big)\nonumber\\&&+\big(M^2-q^2\big)^2 \big(m_j^2-m_W^2+2 M^2-2 q^2\big)-5 m^6\big)  \\
\xi_{e_2}&=& m^6 \big(-m_j^2+m_W^2+3 M^2\big)+m^4 \big(M^2 \big(6 m_j^2-6 m_W^2-7 q^2\big)+3 q^2 \big(m_j^2-m_W^2\big)+2 M^4\big)\nonumber\\&&+m^2 \big(M^4 \big(3
   m_j^2-3 m_W^2+4 q^2\big)+M^2 q^2 \big(-2 m_j^2+2 m_W^2+5 q^2\big)+3 q^4 \big(m_W^2-m_j^2\big)-M^6\big)\nonumber\\&&-\big(M^2-q^2\big) \big(M^4
   \big(8 m_j^2-8 m_W^2-3 q^2\big)-M^2 q^2 \big(3 m_j^2-3 m_W^2+q^2\big)+q^4 \big(m_j^2-m_W^2\big)+4 M^6\big), \\ 
\xi_{e_3}&=& M^2 \big(-2 q^4 \big(m_j^2+5 m^2-m_W^2+2 M^2\big)+q^2 \big(5 m^2-M^2\big) \big(2 m_j^2+m^2-2 m_W^2+M^2\big)\nonumber\\&&+2 \big(m^2-M^2\big)^2 \big(2
   m_j^2+m^2-2 m_W^2+M^2\big)+3 q^6\big),  \\
\xi_{e_4}&=& \big(m_j^2-m_W^2\big) \big((m-M)^2-q^2\big) \big(m^2+3 M^2-q^2\big) \big((m+M)^2-q^2\big), \\
\xi_{e_5}&=& -2 M^2 \big(m^6 m_j^2+m^4 \big(M^2 \big(2 q^2-3 m_j^2\big)+q^2 \big(m_j^2-2 m_W^2+q^2\big)\big)+m^2 \big(M^4 \big(3 m_j^2-q^2\big)+q^2
   \big(q^2 \big(m_j^2-2 m_W^2\big)\nonumber\\&&+3 \big(m_j^2-m_W^2\big){}^2-2 q^4\big)+M^2 q^2 \big(3 q^2-2 m_W^2\big)\big)-\big(M^2-q^2\big)
   \big(M^4 \big(m_j^2+q^2\big)+2 M^2 q^2 \big(m_j^2-2 m_W^2-q^2\big)\nonumber\\&&+q^2 \big(-3 q^2 m_j^2+3 \big(m_j^2-m_W^2\big){}^2+4 q^2
   m_W^2+q^4\big)\big)\big).
\end{eqnarray}

\begin{eqnarray}
\xi_{f_0}&=&-m^2 \big(-m^6-m^4 \big(M^2-3 q^2\big)+m^2 \big(5 M^4+6 M^2 q^2-3 q^4\big)-\big(M^2-q^2\big)^2 \big(3 M^2-q^2\big)\big), \\
\xi_{f_1}&=&m^6 \big(8 m_j^2-8 m_W^2+M^2-7 q^2\big)+m^4 \big(M^2 \big(-3 m_j^2+3 m_W^2-4 q^2\big)+q^2 \big(-11 m_j^2+11 m_W^2+2 q^2\big)-2 M^4\big)\nonumber\\&-&m^2
   \big(M^2-q^2\big) \big(M^2 \big(6 m_j^2-6 m_W^2-4 q^2\big)+q^2 \big(4 m_j^2-4 m_W^2+q^2\big)+3 M^4\big)+\big(M^2-q^2\big)^3  \big(m_j^2-m_W^2\big)+4 m^8,\nonumber \\ & & \\
\xi_{f_2}&=& m^2 \big(m^4 \big(-m_j^2+m_W^2-3 M^2+6 q^2\big)+2 m^2 \big(M^2 \big(-5 m_j^2+5 m_W^2+4 q^2\big)+q^2 \big(m_j^2-m_W^2-3 q^2\big)\big)\nonumber\\&&+M^4
   \big(11 m_j^2-11 m_W^2-2 q^2\big)+M^2 q^2 \big(2 m_j^2-2 m_W^2-5 q^2\big)+q^4 \big(-m_j^2+m_W^2+2 q^2\big)-2 m^6+5 M^6\big), \\ 
\xi_{f_3}&=&m^2 \big(2 q^4 \big(m_j^2+2 m^2-m_W^2+5 M^2\big)+q^2 \big(m^2-5 M^2\big) \big(2 m_j^2+m^2-2 m_W^2+M^2\big)\nonumber\\&&-2 \big(m^2-M^2\big)^2 \big(2
   m_j^2+m^2-2 m_W^2+M^2\big)-3 q^6\big), \\
\xi_{f_4}&=&\big(m_j^2-m_W^2\big) \big(-\big((m-M)^2-q^2\big)\big) \big(3 m^2+M^2-q^2\big) \big((m+M)^2-q^2\big), \\
\xi_{f_5}&=&2 m^2 \big(q^6 \big(m_j^2+3 m^2-2 m_W^2+4 M^2\big)+(m-M)^2 (m+M)^2 \big(m^2 \big(3 m_j^2-2 m_W^2+2 M^2\big)+M^2 \big(5 m_j^2-2 m_W^2\big)\nonumber\\&&+2
   \big(m_j^2-m_W^2\big){}^2\big)-q^4 \big(-2 m_W^2 \big(m_j^2+m^2+4 M^2\big)-m^2 m_j^2+3 M^2 m_j^2+m_j^4+3 m^4+5 m^2 M^2+m_W^4+5
   M^4\big)\nonumber\\&&+q^2 \big(-m^4 \big(5 m_j^2-2 m_W^2+M^2\big)+m^2 \big(-\big(m_j^2-m_W^2\big){}^2-6 M^2 m_W^2+2 M^4\big)+M^2 \big(-M^2 \big(3
   m_j^2+4 m_W^2\big)\nonumber\\&&+5 \big(m_j^2-m_W^2\big){}^2+2 M^4\big)+m^6\big)-q^8\big).
\end{eqnarray}

As far as the $\chi_k$ factors entering the $H_{PV_a}$ functions in eq.~(\ref{NumPasarino1}), they are given as follows 

\begin{eqnarray}
\chi_{k_1}&=&m^2 \big(s_{12} \big(m_i^2+2 m_j^2-3 m_{\ell'}^2-3 m_W^2+2 s_{12}+s_{13}\big)+2 M^2 \big(m_j^2-m_{\ell'}^2-m_W^2\big)\big)\nonumber\\&+&M^2 s_{12}   \big(m_i^2+2 m_j^2-3 m_{\ell'}^2-3 m_W^2+2 s_{12}+s_{13}\big)-s_{12} \big(m_i^2 \big(-2 m_{\ell'}^2+s_{12}+2 s_{13}\big)+s_{12} m_j^2+2 m_{\ell'}^2 \big(m_W^2-s_{12}\big)\nonumber\\&-&\big(s_{12}+s_{13}\big) \big(2 m_W^2-s_{12}\big)\big)+m^4
   \big(-m_j^2+m_{\ell'}^2+m_W^2-s_{12}\big)+M^4 \big(-m_j^2+m_{\ell'}^2+m_W^2-s_{12}\big),\\
\chi_{k_2}&=&-s_{12} \big(-4 m_i^2 m_{\ell'}^2+s_{12} m_i^2-m^2 \big(m_j^2-3 m_{\ell'}^2-m_W^2+s_{12}\big)-M^2 \big(m_j^2-3 m_{\ell'}^2-m_W^2+s_{12}\big)-2   m_j^2 m_{\ell'}^2\nonumber\\&+&s_{12} m_j^2+2 s_{13} m_j^2-4 s_{12} m_{\ell'}^2-2 s_{13} m_{\ell'}^2+6 m_{\ell'}^2 m_W^2+2 m_{\ell'}^4-2 s_{12} m_W^2-2 s_{13}
   m_W^2+s_{12}^2+s_{12} s_{13}\big), \\
\chi_{k_3}&=&-m^2 \big(m_i^2 \big(s_{12}-2 m_{\ell'}^2\big)+M^2 \big(m_j^2+m_{\ell'}^2-m_W^2-s_{12}\big)+s_{13} \big(m_j^2-m_{\ell'}^2-m_W^2+2   s_{12}\big)\nonumber\\&-&m_j^2 m_{\ell'}^2+2 s_{12} m_j^2-4 s_{12} m_{\ell'}^2+3 m_{\ell'}^2 m_W^2+m_{\ell'}^4-3 s_{12} m_W^2+2 s_{12}^2\big)-M^2 \big(2 m_i^2
   m_{\ell'}^2+m_j^2 \big(m_{\ell'}^2+s_{12}-s_{13}\big)\nonumber\\&+&s_{13} \big(m_{\ell'}^2+m_W^2+s_{12}\big)-3 m_{\ell'}^2 m_W^2-m_{\ell'}^4-s_{12}
   m_W^2+s_{12}^2\big)+s_{12} \big(m_i^2 \big(-3 m_{\ell'}^2+s_{12}+s_{13}\big)\nonumber\\&+&m_j^2 \big(-m_{\ell'}^2+s_{12}+s_{13}\big)+\big(2
   m_{\ell'}^2-s_{12}-s_{13}\big) \big(m_{\ell'}^2+2 m_W^2-s_{12}-s_{13}\big)\big)\nonumber\\&+&m^4 \big(m_j^2-m_{\ell'}^2-m_W^2+s_{12}\big)+2 M^4 m_{\ell'}^2,\\  
\chi_{k_4}&=&-m^2 \big(2 m_i^2 m_{\ell'}^2+M^2 \big(m_j^2+m_{\ell'}^2-m_W^2-s_{12}\big)+s_{13} \big(-m_j^2+m_{\ell'}^2+m_W^2+s_{12}\big)+m_j^2 m_{\ell'}^2\nonumber\\&+&s_{12}
   m_j^2-3 m_{\ell'}^2 m_W^2-m_{\ell'}^4-s_{12} m_W^2+s_{12}^2\big)+M^2 \big(m_i^2 \big(2 m_{\ell'}^2-s_{12}\big)+m_j^2 \big(m_{\ell'}^2-2   s_{12}-s_{13}\big)\nonumber\\&+&s_{13} \big(m_{\ell'}^2+m_W^2-2 s_{12}\big)+4 s_{12} m_{\ell'}^2-3 m_{\ell'}^2 m_W^2-m_{\ell'}^4+3 s_{12} m_W^2-2
   s_{12}^2\big)+s_{12} \big(m_i^2 \big(-3 m_{\ell'}^2+s_{12}+s_{13}\big)\nonumber\\&+&m_j^2 \big(-m_{\ell'}^2+s_{12}+s_{13}\big)+\big(2
   m_{\ell'}^2-s_{12}-s_{13}\big) \big(m_{\ell'}^2+2 m_W^2-s_{12}-s_{13}\big)\big)\nonumber\\&+&M^4 \big(m_j^2-m_{\ell'}^2-m_W^2+s_{12}\big)+2 m^4 m_{\ell'}^2 ,\\ 
\chi_{k_5}&=& 2 m^2 \big(s_{12} \big(m_i^2 \big(m_j^2-3 m_{\ell'}^2-m_W^2+s_{12}\big)+m_j^2 \big(-3 m_{\ell'}^2-3 m_W^2+2 s_{12}+s_{13}\big)+m_j^4-3
   s_{12} m_{\ell'}^2-s_{13} m_{\ell'}^2\nonumber\\&+&4 m_{\ell'}^2 m_W^2+2 m_{\ell'}^4-3 s_{12} m_W^2-3 s_{13} m_W^2+2 m_W^4+s_{12}^2+s_{12} s_{13}\big)+M^2 \big(-2
   m_j^2 \big(m_{\ell'}^2+m_W^2\big)+m_j^4\nonumber\\&+&2 s_{12} \big(m_{\ell'}^2+m_W^2\big)+\big(m_{\ell'}^2-m_W^2\big){}^2-s_{12}^2\big)\big)+2 M^2
   s_{12} \big(m_i^2 \big(m_j^2-3 m_{\ell'}^2-m_W^2+s_{12}\big)\nonumber\\&+&m_j^2 \big(-3 m_{\ell'}^2-3 m_W^2+2 s_{12}+s_{13}\big)+m_j^4-3 s_{12}
   m_{\ell'}^2-s_{13} m_{\ell'}^2+4 m_{\ell'}^2 m_W^2+2 m_{\ell'}^4-3 s_{12} m_W^2-3 s_{13} m_W^2\nonumber\\&+&2 m_W^4+s_{12}^2+s_{12} s_{13}\big)-s_{12} \big(2 m_i^2  \big(m_j^2 \big(-2 m_{\ell'}^2+s_{12}+2 s_{13}\big)+m_{\ell'}^2 \big(6 m_W^2-4 s_{12}-2 s_{13}\big)+2 m_{\ell'}^4\nonumber\\&-&\big(s_{12}+s_{13}\big)
   \big(2 m_W^2-s_{12}\big)\big)+m_i^4 \big(s_{12}-4 m_{\ell'}^2\big)+2 m_j^2 \big(2 m_{\ell'}^2
   \big(m_W^2-s_{12}\big)-\big(s_{12}+s_{13}\big) \big(2 m_W^2-s_{12}\big)\big)+s_{12} m_j^4\nonumber\\&-&\big(2   m_{\ell'}^2-s_{12}-s_{13}\big) \big(m_{\ell'}^2 \big(4 m_W^2-2 s_{12}\big)-4 \big(s_{12}+s_{13}\big) m_W^2+4 m_W^4+s_{12}   \big(s_{12}+s_{13}\big)\big)\big)\nonumber\\&+&m^4 \big(-\big(m_j^2-\big(m_{\ell'}-m_W\big){}^2+s_{12}\big)\big)   \big(m_j^2-\big(m_{\ell'}+m_W\big){}^2+s_{12}\big)\nonumber\\&-&M^4 \big(m_j^2-\big(m_{\ell'}-m_W\big){}^2+s_{12}\big)   \big(m_j^2-\big(m_{\ell'}+m_W\big){}^2+s_{12}\big)  
\end{eqnarray}

\section{Some useful integrals}\label{Int-y}

As we mentioned in the text, analytical expressions for the double integrals in eqs. (\ref{IntFeyn1}), (\ref{IntFeyn2}) and (\ref{IntFeyn3}) are not easy to obtain. However, the first integrals over the $y$-Feynman parameter can be derived from the following expressions

\begin{equation}
\int_0^{1-x}\frac{dy}{D_j}=-\frac{2}{\Lambda}\left(T_{+}-T_{-}\right),
\end{equation}

\begin{equation}
\int_0^{1-x}\frac{y dy}{D_j}=\frac{\left(T_{+}-T_{-}\right)}{q^2\Lambda}\left(x(M^2-m^2)+q^2(x-1)\right)+\frac{\left(\theta_m-\theta_M  \right)}{2q^2},
\end{equation}

\begin{eqnarray}
\int_0^{1-x}\frac{y^2 dy}{D_j}&=&\frac{\left(T_--T_+\right)}{\Lambda  q^4} \left(2 q^2 (x-1) m_j^2+x^2 \left(m^2-M^2\right)^2-2 q^2 x \left(m^2 (x-1)+m_W^2\right)+q^4 (x-1)^2\right)\nonumber\\ & &-\frac{\left(\theta _m-\theta _M\right)}{2 q^4} \left(x \left(M^2-m^2\right)+q^2 (x-1)\right)+\frac{1-x}{q^2},
\end{eqnarray}

\begin{eqnarray}
\int_0^{1-x}\ln(D_j)dy&=&\frac{\Lambda \left(T_--T_+\right)}{q^2}+\frac{\left(\theta _m-\theta _M\right) \left(x \left(M^2-m^2\right)+q^2 (x-1)\right)}{2 q^2}\nonumber\\ & & -(x-1) \left(\log \left(x \left(m^2 (x-1)+m_W^2\right)-(x-1) m_j^2\right)-2\right),
\end{eqnarray}
where we have defined the functions that follow

\begin{equation}
\Lambda=\sqrt{-4 q^2 (x-1) m_j^2+2 q^2 x \left(m^2 (x-1)+2 m_W^2+M^2 (x-1)\right)-x^2 \left(m^2-M^2\right)^2+q^4 \left(-(x-1)^2\right)},
\end{equation}

\begin{equation}
T_+=\tan ^{-1}\left(\frac{x \left(M^2-m^2\right)+q^2 (x-1)}{\Lambda }\right),\quad T_-=\tan ^{-1}\left(\frac{x \left(M^2-m^2\right)-q^2 (x-1)}{\Lambda }\right),
\end{equation}

\begin{equation}
\theta_M=\log \left((x-1) m_j^2-x \left(m_W^2+M^2 (x-1)\right)\right),\quad 
\theta_m=\log \left((x-1) m_j^2-x \left(m_W^2+m^2 (x-1)\right)\right).
\end{equation}

\section{Kinematics for the $L^{-}(P, M)\to\ell^{-}(p,m) \ell^{\prime -}(p_1, m_{\ell'}) \ell^{\prime +}(p_2 , m_{\ell'})$ decays} \label{Kinematics}

Because of the necessity of antisymmetrizing the amplitude when $\ell=\ell^\prime$, the total contribution for the sum of the penguin and box diagrams in this case is given by 

\begin{eqnarray}
\mathcal{M}_{\ell=\ell'}&=&iG_F^2 m_W^2\left(-\frac{\beta_{F_a}}{4}+8m_W^2\beta_{H_a}\right)\,\bar{u}(p)\gamma_{\lambda}(1- \gamma_5) u(P) \times \bar{u}(p_{1})\gamma^{\lambda}(1- \gamma_5) v(p_2)-(p\leftrightarrow p_1)\nonumber\\ & +&i G_F^2 m_W^2 s_W^2\beta_{F_a}\,\bar{u}(p)\gamma_{\lambda}(1- \gamma_5) u(P) \times \bar{u}(p_{1})\gamma^{\lambda} v(p_2)-(p\leftrightarrow p_1),\label{Amp-ll}
\end{eqnarray}

On the other hand, when $\ell\neq\ell'$, there is only one penguin diagram since the neutral $Z$ boson does not change flavor. Besides, we have to add the box diagram interchanging $\ell^-(p)\leftrightarrow\ell^{
\prime-}(p_1)$. Therefore, we have  

\begin{eqnarray}
\mathcal{M}_{\ell\neq\ell'}&=&iG_F^2 m_W^2\left(-\frac{\beta_{F_a}}{4}+8m_W^2\beta_{H_a}\right)\,\bar{u}(p)\gamma_{\lambda}(1- \gamma_5) u(P) \times \bar{u}(p_{1})\gamma^{\lambda}(1- \gamma_5) v(p_2)\nonumber\\ & +&i G_F^2 m_W^2 s_W^2\beta_{F_a}\,\bar{u}(p)\gamma_{\lambda}(1- \gamma_5)u(P) \times \bar{u}(p_{1})\gamma^{\lambda} v(p_2)\nonumber\\ &+& 8iG_F^2 m_W^4\hat{\beta}_{H_a}\bar{u}(p_1)\gamma_{\lambda}(1-\gamma_5) u(P) \times \bar{u}(p)\gamma^{\lambda}(1- \gamma_5) v(p_2)\label{Amp-llp}
\end{eqnarray}
where $\beta_{H_a}$ has been defined in the main text and

\begin{equation}
\hat{\beta}_{H_a}=\sum_{j,i} U_{Lj}U^*_{\ell i}  U_{\ell' j}U^*_{\ell' i} H_a(m_i^2, m_j^2).
\end{equation}

In the Petcov's approximation, taking only the dominant term and since the contribution of the penguin and box diagrams have opposite sign, the dominant terms are given by the second terms in eqs. (\ref{Amp-ll}) and (\ref{Amp-llp}), respectively. Therefore, $|\mathcal{M}^2|$ is given by    
\begin{eqnarray}
|\mathcal{M}^2|& = & \frac{G_F^4 s_W^4}{4 \pi^4}\left(\sum_j U_{Lj}U^*_{lj}m_j^2 \log\left(\frac{m_W^2}{m_j^2}\right) \right)^2 T_s,
 \label{M2}
\end{eqnarray}
where 

\begin{equation}
T_s=-16 \left(-4 s_{13} m_{\ell'}^2+2 m_{\ell'}^4-s_{12} \left(m^2+M^2-2 s_{13}\right)+2 \left(m^2-s_{13}\right) \left(M^2-s_{13}\right)+s_{12}^2\right)
\end{equation}

for $\ell\neq\ell'$, and 

\begin{eqnarray}
T_s&=&-16 \big(-s_{13}\big(5 m^2+M^2\big)+s_{13}^2+2 \big(9 m^4-2 s_{13} \big(4
   m^2+M^2\big)\nonumber\\&+&s_{12} \big(-3 m^2-M^2+s_{13}\big)+5 m^2 M^2+s_{12}^2+2 s_{13}^2\big)\big)
\end{eqnarray}

 when $\ell=\ell'$ ($m=m_{\ell'}$).

The unpolarized differential decay width for the $L^{-}(P)\to\ell^{-}(p) \ell^{\prime -}(p_1) \ell^{\prime +}(p_2)$  decays is given by 

\begin{equation}
\Gamma=\frac{1/N}{4(4 \pi)^3 M^3}\int |\mathcal{M}|^2 ds_{12}ds_{13},\label{dif-witdh}
\end{equation}
where $N$ is the number of identical particles in the final state and $s_{12}=(p_1+p_2)^2=q^2$ and $s_{13}=(p_1+p)^2$. The corresponding integration limits are given by
\begin{equation}
s_{13}^{\pm}=\frac{(s_{12})(M^2-s_{12}-m^2)}{2s_{12}}+m_{\ell'}^2+m^2
\pm\frac{\sqrt{\lambda(M^2, s_{12}, m^2)\lambda(s_{12}, m_{\ell'}^2,m_{\ell'}^2)}}{2s_{12}}
.
\end{equation}
and
\begin{equation}
4 m_{\ell'}^2\leq s_{12}\leq (M-m)^2\,.
\end{equation}

\section{Fits for $Z\tau \mu$, $Z\tau e$ and $Z\mu e$ effective vertices} \label{AdditionalFits}

The numerical values for the $Q_k$ and $R_k$ factors involved in of our fits for the $Z\tau \mu$, $Z\tau e$ and $Z\mu e$ effective vertices are given as follows

\begin{table}[!htbp]
\begin{center}
\begin{tabular}{|c|c|c||c|c|}
\multicolumn{1}{c}{$Z\tau\mu$ $(q^{2}=4m_{\mu}^{2})$} & \multicolumn{1}{c}{$Q_{R_{k}}$} & \multicolumn{1}{c}{$R_{R_{k}}$} & \multicolumn{1}{c}{$Q_{I_{k}}$} & \multicolumn{1}{c}{$R_{I_{k}}$}\tabularnewline
\hline 
\hline 
$a$ & $4.63706$ & $11.5451$ & $-7.14896\times10^{-6}$ & $3.4098$\tabularnewline
\hline 
$b$ & $1.38093\times10^{-5}$ & $-3.31777\times10^{-4}$ & $9.85094\times10^{-11}$ & $-6.76208\times10^{-5}$\tabularnewline
\hline 
$c$ & $-1.49047\times10^{-5}$ & $3.62348\times10^{-3}$ & $-7.884\times10^{-10}$ & $5.4035\times10^{-4}$\tabularnewline
\hline 
$d$ & $-9.20638\times10^{-6}$ & $1.2469\times10^{-4}$ & $-4.9267\times10^{-11}$ & $3.38191\times10^{-5}$\tabularnewline
\hline 
$e$ & $2.04592\times10^{-3}$ & $191.959$ & $4.69628\times10^{-4}$ & $-126.096$\tabularnewline
\hline 
$f$ & $-1.26365\times10^{-5}$ & $-11.8554$ & $-2.95163\times10^{-5}$ & $8.05527$\tabularnewline
\hline 
\end{tabular}
\par\end{center}
\caption{Values for the $Q_{R_k}$ ($Q_{I_k}$) and $R_{R_k}$ ($R_{I_k}$) coefficients of the $Z\tau\mu$ vertex for $q^2=4m_\mu^2$.}\label{Table-Ztau-mu-qmu}
\end{table}

\medskip{}

\begin{table}[!htbp]
\begin{center}
\begin{tabular}{|c|c|c||c|c|}
\multicolumn{1}{c}{$Z\tau\mu$ $(q^{2}=4m_{e}^{2})$} & \multicolumn{1}{c}{$Q_{R_{k}}$} & \multicolumn{1}{c}{$R_{R_{k}}$} & \multicolumn{1}{c}{$Q_{I_{k}}$} & \multicolumn{1}{c}{$R_{I_{k}}$}\tabularnewline
\hline 
\hline 
$a$ & $4.63709$ & $22.2936$ & $-1.6966\times10^{-10}$ & $3.40516$\tabularnewline
\hline 
$b$ & $1.3809\times10^{-5}$ & $-6.24913\times10^{-4}$ & $2.31697\times10^{-15}$ & $-6.71321\times10^{-5}$\tabularnewline
\hline 
$c$ & $-1.49044\times10^{-4}$ & $-3.92512\times10^{-2}$ & $-1.89825\times10^{-14}$ & $6.2734\times10^{-4}$\tabularnewline
\hline 
$d$ & $-9.20617\times10^{-6}$ & $-0.191951$ & $-1.0909\times10^{-15}$ & $2.6232\times10^{-5}$\tabularnewline
\hline 
$e$ & $3.63186\times10^{-3}$ & $8.17424\times10^{6}$ & $4.74754\times10^{-4}$ & $-5.3963\times10^{6}$\tabularnewline
\hline 
$f$ & $-2.2432\times10^{-4}$ & $-504881$ & $-2.93231\times10^{-5}$ & $333301$\tabularnewline
\hline 
\end{tabular}
\par\end{center}
\caption{Same as Table \ref{Table-Ztau-mu-qmu} but considering $q^2=4m_e^2$.}\label{Table-Ztau-mu-qe}
\end{table}

\begin{table}[!ht]
\begin{center}
\begin{tabular}{|c|c|c||c|c|}
\multicolumn{1}{c}{$Z\tau e$ $(q^{2}=4m_{\mu}^{2})$} & \multicolumn{1}{c}{$Q_{R_{k}}$} & \multicolumn{1}{c}{$R_{R_{k}}$} & \multicolumn{1}{c}{$Q_{I_{k}}$} & \multicolumn{1}{c}{$R_{I_{k}}$}\tabularnewline
\hline 
\hline 
$a$ & $4.63706$ & $11.5451$ & $-7.14896\times10^{-6}$ & $3.4098$\tabularnewline
\hline 
$b$ & $6.72054\times10^{-8}$ & $-1.61465\times10^{-6}$ & $4.79412\times10^{-13}$ & $-3.29087\times10^{-7}$\tabularnewline
\hline 
$c$ & $-1.49047\times10^{-4}$ & $3.61659\times10^{-3}$ & $-7.88464\times10^{-10}$ & $5.4042\times10^{-4}$\tabularnewline
\hline 
$d$ & $-3.86832\times10^{-8}$ & $-5.66645\times10^{-3}$ & $-2.39753\times10^{-13}$ & $1.64583\times10^{-7}$\tabularnewline
\hline 
$e$ & $2.04592\times10^{-3}$ & $191.962$ & $4.69628\times10^{-4}$ & $-126.095$\tabularnewline
\hline 
$f$ & $-6.09267\times10^{-7}$ & $-5.92939\times10^{-2}$ & $-1.43646\times10^{-7}$ & $3.920023\times10^{-2}$\tabularnewline
\hline 
\end{tabular}
\par\end{center}

\caption{Values for the $Q_{R_k}$ ($Q_{I_k}$) and $R_{R_k}$ ($R_{I_k}$) coefficients of the $Z\tau e$ vertex for $q^2=4m_\mu^2$.}\label{Table-Ztau-e-qmu}
\end{table}

\medskip{}

\begin{table}[!ht]
\begin{center}
\begin{tabular}{|c|c|c||c|c|}
\multicolumn{1}{c}{$Z\tau e$ $(q^{2}=4m_{e}^{2})$} & \multicolumn{1}{c}{$Q_{R_{k}}$} & \multicolumn{1}{c}{$R_{R_{k}}$} & \multicolumn{1}{c}{$Q_{I_{k}}$} & \multicolumn{1}{c}{$R_{I_{k}}$}\tabularnewline
\hline 
\hline 
$a$ & $4.63709$ & $22.2262$ & $-1.6966\times10^{-10}$ & $3.40516$\tabularnewline
\hline 
$b$ & $6.72036\times10^{-8}$ & $-3.05754\times10^{-4}$ & $1.12759\times10^{-17}$ & $-3.26709\times10^{-7}$\tabularnewline
\hline 
$c$ & $-1.49043\times10^{-4}$ & $-.107576$ & $-1.89741\times10^{-14}$ & $6.26186\times10^{-4}$\tabularnewline
\hline 
$d$ & $-4.95278\times10^{-8}$ & $19.097$ & $-5.68205\times10^{-18}$ & $1.64383\times10^{-7}$\tabularnewline
\hline 
$e$ & $3.63189\times10^{-3}$ & $8.17296\times10^{6}$ & $4.74754\times10^{-4}$ & $-5.3963\times10^{6}$\tabularnewline
\hline 
$f$ & $-1.08821\times10^{-6}$ & $-2468.32$ & $-1.42705\times10^{-7}$ & $1622.06$\tabularnewline
\hline 
\end{tabular}
\par\end{center}
\caption{Same as Table \ref{Table-Ztau-e-qmu} but considering $q^2=4m_e^2$.}\label{Table-Ztau-e-qe}
\end{table}

\begin{table}[!ht]
\begin{center}
\begin{tabular}{|c|c|c||c|c|}
\multicolumn{1}{c}{$Z\mu e$ $(q^{2}=4m_{e}^{2})$} & \multicolumn{1}{c}{$Q_{R_{k}}$} & \multicolumn{1}{c}{$R_{R_{k}}$} & \multicolumn{1}{c}{$Q_{I_{k}}$} & \multicolumn{1}{c}{$R_{I_{k}}$}\tabularnewline
\hline 
\hline 
$a$ & $4.63701$ & $31.6578$ & $-1.55723\times10^{-10}$ & $1.15008$\tabularnewline
\hline 
$b$ & $4.15019\times10^{-9}$ & $-1.01036\times10^{-7}$ & $7.02341\times10^{-19}$ & $-2.03165\times10^{-8}$\tabularnewline
\hline 
$c$ & $-5.6794\times10^{-7}$ & $-2.40973$ & $-1.60743\times10^{-13}$ & $2.42044\times10^{-3}$\tabularnewline
\hline 
$d$ & $-9.81338\times10^{-8}$ & $2359.07$ & $-1.54223\times10^{-16}$ & $-2.54211\times10^{-6}$\tabularnewline
\hline 
$e$ & $1.37244\times10^{-5}$ & $32441.2$ & $1.78855\times10^{-6}$ & $-20427.4$\tabularnewline
\hline 
$f$ & $-8.46878\times10^{-8}$ & $-2024.81$ & $-8.70909\times10^{-9}$ & $99.6226$\tabularnewline
\hline 
\end{tabular}
\par\end{center}

\caption{Values for the $Q_{R_k}$ ($Q_{I_k}$) and $R_{R_k}$ ($R_{I_k}$) coefficients of the $Z\mu e$ vertex for $q^2=4m_\mu^2$.}\label{Table-Zmu-e-qmu}
\end{table}

\clearpage

\section{Expansion of the PaVe functions around $m_j^2=0$} \label{Expansion}


The scalar PaVe functions involve in eq. (\ref{NumPasarino})  calculation are defined as follows

\begin{equation}
B_0(p^2, m_j^2, m_W^2)=(i\pi^2)^{-1}\int \frac{d^nk}{\left(k^2-m_j^2\right)\left[\left(k+p\right)^2-m_W^2\right]},\label{B1}
\end{equation}

\begin{equation}
B_0(q^2, m_j^2, m_j^2)=(i\pi^2)^{-1}\int \frac{d^nk}{\left(k^2-m_j^2\right)\left[\left(k+q\right)^2-m_j^2\right]},\label{B3}
\end{equation}

\begin{equation}
B_0(0, m_j^2, m_W^2)=(i\pi^2)^{-1}\int \frac{d^nk}{\left(k^2-m_j^2\right)\left(k^2-m_W^2\right)},\label{B4}
\end{equation}

\begin{equation}
C_0(p^2, P^2, q^2,m_j^2, m_W^2, m_j^2)=(i\pi^2)^{-1}\int \frac{d^nk}{\left(k^2-m_W^2\right)\left[\left(k+p\right)^2-m_j^2\right]\left[\left(k+P\right)^2-m_j^2\right]}, \label{C1}
\end{equation}where $p^2=m^2$, $P^2=M^2$ and $q^2=(P-p)^2=m^2+M^2-2P \cdot p$.

If we do an expansion around $m_j^2=0$, for the equations (\ref{B1}, \ref{B3}, \ref{B4}, \ref{C1}) following the same strategy that Cheng and Li for the $\mu\to e\gamma$ decay \cite{LibroCheng}, we have that

\begin{eqnarray}
B_0(p^2, m_j^2, m_W^2)&\approx & B_0(p^2, 0, m_W^2)+m_j^2C_0(0,p^2, p^2, 0, 0,m_W^2) + \vartheta (m_j^4),\label{EB1}
\end{eqnarray}

\begin{eqnarray}
B_0(q^2, m_j^2, m_j^2)&\approx &  B_0(q^2, 0, 0)+2m_j^2 C_0(0,q^2, q^2, 0, 0,0) + \vartheta (m_j^4),\label{EB3}
\end{eqnarray}

\begin{eqnarray}
B_0(0, m_j^2, m_W^2)&\approx & B_0(0, 0, m_W^2)+m_j^2 \frac{A_0(m_W^2)}{m_W^4} + \vartheta (m_j^4),\label{EB4}
\end{eqnarray}

\begin{eqnarray}
C_0(p^2, P^2, q^2,m_j^2, m_W^2, m_j^2)&\approx & C_0(p^2, P^2, q^2,0, m_W^2, 0)+m_j^2\left[D_0(p^2,0,q^2,P^2,p^2,q^2,m_W^2,0,0,0)\right.\nonumber\\&+&\left. D_0(p^2,q^2,0,P^2,P^2,q^2,m_W^2,0,0,0) \right]+ \vartheta (m_j^4)\label{EC1}
\end{eqnarray}

Now, with the help of the Package-X program, we can obtain analytical expressions for the next functions

\begin{equation}
B_0(p^2, 0, m_W^2)=\Delta+\frac{\left(p^2-m_W^2\right)}{p^2}\log \left(\frac{m_W^2}{m_W^2-p^2}\right)+\log \left(\frac{\mu ^2}{m_W^2}\right)+2,\label{B0p20mW}
\end{equation}

\begin{equation}
B_0(q^2, 0, 0)=\Delta+\log \left(-\frac{\mu ^2}{q^2}\right)+2,\label{B0q200}
\end{equation}

\begin{equation}
B_0(0, 0, m_W^2)=\Delta+\log \left(\frac{\mu ^2}{m_W^2}\right)+1,\label{B000mW}
\end{equation}

\begin{equation}
C_0(0, q^2,q^2,0,0,0)=-\frac{\Delta_I+\log \left(-\frac{\mu ^2}{q^2}\right)}{q^2},\label{C00q2q2000}
\end{equation}
with $\Delta_I\sim\Delta$ but associated with an infrared divergence.

\begin{equation}
C_0(0, p^2,p^2,0,0,m_W^2)=\frac{\Delta_I+\log \left(\frac{\mu ^2}{m_W^2}\right)}{m_W^2-p^2}+\frac{\left(m_W^2+p^2\right)}{p^2 \left(m_W^2-p^2\right)}\log  \left(\frac{m_W^2}{m_W^2-p^2}\right),\label{C00p2p20mW}
\end{equation}

\begin{eqnarray}
D_0(p^2,0,q^2,P^2,p^2,q^2,m_W^2,0,0,0)&=& \frac{1}{q^2}\left[\frac{\Delta_I+\log \left(\frac{\mu ^2}{m_W^2}\right)}{m_W^2-p^2}+\frac{\left(m_W^2-P^2\right) \log
   \left(-\frac{m_W^2}{q^2}\right)}{-m_W^2 \left(p^2+P^2-q^2\right)+m_W^4+p^2 P^2}\right. \nonumber\\ &-& \left.\frac{\left(m_W^2-P^2\right) \log
   \left(\frac{m_W^2}{m_W^2-P^2}\right)}{-m_W^2 \left(p^2+P^2-q^2\right)+m_W^4+p^2 P^2}\right. \\&+&\left.\frac{\log
   \left(\frac{m_W^2}{m_W^2-p^2}\right) \left(-m_W^2 \left(p^2+P^2-2 q^2\right)+m_W^4+p^2 P^2\right)}{\left(m_W^2-p^2\right)
   \left(-m_W^2 \left(p^2+P^2-q^2\right)+m_W^4+p^2 P^2\right)}\right]\nonumber,\label{D0p20q2P2p2mW000}
\end{eqnarray}

\begin{eqnarray}
D_0(p^2,q^2,0,P^2,P^2,q^2,m_W^2,0,0,0)&=&\frac{1}{q^2}\left[\frac{\Delta_I+\log \left(\frac{\mu ^2}{m_W^2}\right)}{m_W^2-P^2}+\frac{\left(m_W^2-p^2\right) \log
   \left(-\frac{m_W^2}{q^2}\right)}{-m_W^2 \left(p^2+P^2-q^2\right)+m_W^4+p^2 P^2}\right.\nonumber\\& -&\left.\frac{\left(m_W^2-p^2\right) \log
   \left(\frac{m_W^2}{m_W^2-p^2}\right)}{-m_W^2 \left(p^2+P^2-q^2\right)+m_W^4+p^2 P^2}\right.\\&+&\left.\frac{\log
   \left(\frac{m_W^2}{m_W^2-P^2}\right) \left(-m_W^2 \left(p^2+P^2-2 q^2\right)+m_W^4+p^2 P^2\right)}{\left(m_W^2-P^2\right)
   \left(-m_W^2 \left(p^2+P^2-q^2\right)+m_W^4+p^2 P^2\right)}\right]\nonumber.
   \label{D0p2q2P2P2q2mW000}
\end{eqnarray}

Replacing eqs. (\ref{EB1}, \ref{EB3}, \ref{EB4}, \ref{EC1}) and subsequently eqs. (\ref{B000mW}, \ref{B0p20mW}, \ref{B0q200}, \ref{C00p2p20mW}, \ref{C00q2q2000}, \ref{D0p20q2P2p2mW000} and \ref{D0p2q2P2P2q2mW000}) into (\ref{NumPasarino}) we obtain eq. (\ref{Fa-expanded}), with the $f_Q$
and $f_R$ factors given as follows

\begin{eqnarray}
f_{Q_{a_1}}&=& -q^2 \left(-m^2-m M+m_W^2-M^2+q^2\right) \left(-m^2+m M+m_W^2-M^2+q^2\right),\\
f_{Q_{a_2}}&=&\frac{1}{2 m^2}\left((-\left(m^2-m_W^2\right) \left(m^4-m_W^2 \left(m^2-M^2+q^2\right)+m^2 \left(M^2+q^2\right)-2 \left(M^2-q^2\right)^2\right)\right), \\
f_{Q_{a_3}}&=&\frac{1}{2 M^2}\left(-\left(M^2-m_W^2\right) \left(-2 m^4+m_W^2 \left(m^2-M^2-q^2\right)+m^2 \left(M^2+4 q^2\right)+M^4+M^2 q^2-2 q^4\right)\right),\nonumber\\ & & \\
f_{Q_{a_4}}&=&\frac{1}{2}\left(q^2 \left(3 \left(m^2+M^2-q^2\right)-2 m_W^2\right)\right), \\
f_{Q_{a_5}}&=& \frac{1}{2} \left(\lambda(m^2,M^2,q^2) \log \left(\frac{\mu ^2}{m_W^2}\right)+i \pi  q^2 \left(3 \left(m^2+M^2-q^2\right)-2 m_W^2\right)\right).
\end{eqnarray}

\begin{eqnarray}
f_{R_{a_1}}&=& 2 q^2 \left(-m^2+m_W^2-M^2+q^2\right), \\
f_{R_{a_2}}&=& \frac{1}{m^2 \alpha}\left( -m^8+2 m^6 \left(m_W^2+q^2\right)+m^4 \left(-2 m_W^4+M^4-q^4\right)\right.\nonumber\\&+& \left. m^2 m_W^2 \left(-2 q^2 m_W^2+m_W^4-M^4+ 4 M^2 q^2-3
   q^4\right)-m_W^2 \left(M^2-q^2\right) \left(m_W^2-M^2+q^2\right){}^2 \right),\\
f_{R_{a_3}}&=& \frac{1}{M^2 \alpha}\left( -M^4 \left(\left(M^2-q^2\right)^2-m^4\right)+m_W^6 \left(-m^2+M^2+q^2\right)+2 m_W^4 \left(m^4-2 m^2 q^2-M^4-M^2 q^2+q^4\right)\right.\nonumber\\&+& \left. m_W^2
   \left(-m^6-m^4 \left(M^2-3 q^2\right)+m^2 \left(4 M^2 q^2-3 q^4\right)+ 2 M^6-3 M^2 q^4+q^6\right) \right), \\
f_{R_{a_4}}&=& \frac{1}{\alpha}\left(m^6-m^4 \left(m_W^2+M^2+2 q^2\right)+m^2 \left(2 M^2 m_W^2+q^2 \left(q^2-m_W^2\right)-M^4\right)- M^4 \left(m_W^2+2 q^2\right)\right.\nonumber\\ &+& \left.M^2 q^2
   \left(q^2-m_W^2\right)+2 q^2 m_W^2 \left(m_W^2+q^2\right)+M^6 \right), \\
f_{R_{a_5}}&=& \frac{1}{\alpha}\big( i \pi  \big(m^6-m^4 \big(M^2+2 q^2\big)+m^2 \big(q^4-M^4\big)-m_W^2 \big(m^4+m^2 \big(q^2-2 M^2\big)+M^4+M^2 q^2-2 q^4\big)\nonumber\\ &+&2 q^2 m_W^4+\big(M^3-M q^2\big)^2\big) \big),
\end{eqnarray}

and $\alpha=m^2 \left(M^2-m_W^2\right)+m_W^2 \left(m_W^2-M^2+q^2\right)$.\\

Something remarkable at this point is:

The factor $Q_a$ has an ultraviolet divergence $\Delta$, as it can be seen in eq. (\ref{Qa}), but this divergence is independent of the neutrino mass. Then this divergence will vanish when we sum over the three families (GIM-mechanism), as it was mentioned previously.

Although there are infrared divergences $\Delta_I$ on the eqs. (\ref{C00q2q2000}, \ref{C00p2p20mW}, \ref{D0p20q2P2p2mW000}, \ref{D0p2q2P2P2q2mW000}),  the factor $R_a$ is free of them. Further, there is no dependence on the renormalization scale, and these results are in agreement with our numerical fits. Taking into account the imaginary part of the $C_0(m^2, M^2, q^2,0,m_W^2,0)$ function, it is possible to derive analytically that the imaginary parts appearing in the last column of Table 9 are exactly $\pi$.

\begin{table}[!h]
\begin{center}
\begin{tabular}{|l|l|l|}
\hline
    Vertex   & $R_a$ (Numerical Fits)     & $R_a$ (eq. \ref{R_a})       \\ \hline
$Z\tau\mu$ ($q^2_{min}=4m_\mu^2$ for $\tau^-\to\mu^-\mu^-\mu^+$)  & 11.5451+i3.4098  & 11.3949+i3.14159 \\ \hline
$Z\tau\mu$ ($q^2_{min}=4m_e^2$ for $\tau^-\to\mu^-e^-e^+$)   & 22.2936+i3.40516 & 22.0456+i3.14159  \\ \hline
$Z\tau e$ ($q^2_{min}=4m_\mu^2$ for $\tau^-\to e^-\mu^-\mu^+$) & 11.5451+i3.4098  & 11.3976+i3.14159  \\ \hline
$Z\tau e$ ($q^2_{min}=4m_e^2$ for $\tau^-\to e^-e^-e^+$)   & 22.2262+i3.40516 & 22.0483+i3.14159  \\ \hline
$Z\mu e$ ($q^2_{min}=4m_e^2$  for $\mu^-\to e^-e^-e^+$)    & 31.6578+i1.15008 & 22.7478+i3.14159  \\ \hline

\end{tabular}
\caption{Comparison for the $R_a$ factor using numerical fits vs eq. (\ref{R_a}).}\label{Compariso Ra}
\end{center}
\end{table}


\addcontentsline{toc}{section}{References}


\begin{thebibliography}{9}
\bibitem{SM}
  S.~L.~Glashow,
  Nucl.\ Phys.\  {\bf 22} (1961) 579;
  S.~Weinberg,
  Phys.\ Rev.\ Lett.\  {\bf 19} (1967) 1264;
  A.~Salam,
  Conf.\ Proc.\ C {\bf 680519} (1968) 367.
  
\bibitem{nuosc}
  Y.~Fukuda {\it et al.} [Super-Kamiokande Collaboration],
  Phys.\ Rev.\ Lett.\  {\bf 81}, 1562 (1998);
%
  Q.~R.~Ahmad {\it et al.} [SNO Collaboration],
  Phys.\ Rev.\ Lett.\  {\bf 87}, 071301 (2001);
%
{\bf 89}, 011301 (2002).
  
  \bibitem{CKM}
  N.~Cabibbo,
  Phys.\ Rev.\ Lett.\  {\bf 10} (1963) 531; 
  M.~Kobayashi and T.~Maskawa,
  Prog.\ Theor.\ Phys.\  {\bf 49} (1973) 652.
  
  \bibitem{PMNS}
  B.~Pontecorvo,
  Sov.\ Phys.\ JETP {\bf 10} (1960) 1236
   [Zh.\ Eksp.\ Teor.\ Fiz.\  {\bf 37} (1959) 1751];
  Z.~Maki, M.~Nakagawa and S.~Sakata,
  Prog.\ Theor.\ Phys.\  {\bf 28}, 870 (1962).
  
\bibitem{Glashow:1970gm} 
  S.~L.~Glashow, J.~Iliopoulos and L.~Maiani,
  Phys.\ Rev.\ D {\bf 2}, 1285 (1970).

 \bibitem{LibroCheng}
 T. P. Cheng and L. F. Li, “Gauge Theory Of Elementary Particle Physics,” Oxford,
Uk: Clarendon (1984) 536 P. (Oxford Science Publications) 

\bibitem{Calibbi:2017uvl} 
  L.~Calibbi and G.~Signorelli,
  Riv.\ Nuovo Cim.\  {\bf 41},  1 (2018)
  
\bibitem{Petcov:1976ff} 
  S.~T.~Petcov,
  Sov.\ J.\ Nucl.\ Phys.\  {\bf 25}, 340 (1977)
  [Yad.\ Fiz.\  {\bf 25}, 641 (1977)]
  Erratum: [Sov.\ J.\ Nucl.\ Phys.\  {\bf 25}, 698 (1977)]
  Erratum: [Yad.\ Fiz.\  {\bf 25}, 1336 (1977)].

\bibitem{Lee:1977tib} 
  B.~W.~Lee and R.~E.~Shrock,
  Phys.\ Rev.\ D {\bf 16}, 1444 (1977).

\bibitem{Pham:1998fq} 
  X.~Y.~Pham,
  Eur.\ Phys.\ J.\ C {\bf 8}, 513 (1999).
  
\bibitem{Patrignani:2016xqp} 
  C.~Patrignani {\it et al.} [Particle Data Group],
  Chin.\ Phys.\ C {\bf 40},  100001 (2016).
  
 \bibitem{Referencia}
 Sw. Banerjee et al [HFLAV-Tau group], available at http://www.slac.stanford.edu/xorg/hflav/tau/spring-2017/tau-report-web.pdf 
  
  \bibitem{Fits}
  I.~Esteban, M.~C.~Gonz\'alez-Garcia, M.~Maltoni, I.~Mart\'inez-Soler and T.~Schwetz,
  JHEP {\bf 1701}, 087 (2017);
%
  F.~Capozzi, E.~Di Valentino, E.~Lisi, A.~Marrone, A.~Melchiorri and A.~Palazzo,
  Phys.\ Rev.\ D {\bf 95}, 096014 (2017);
  P.~F.~de Salas, D.~V.~Forero, C.~A.~Ternes, M.~Tortola and J.~W.~F.~Valle,
  Phys.\ Lett.\ B {\bf 782}, 633 (2018)
  
\bibitem{Kou:2018nap} 
  E.~Kou {\it et al.},
  arXiv:1808.10567 [hep-ex].
  
  \bibitem{KLN}
  T.~Kinoshita,
  J.\ Math.\ Phys.\  {\bf 3} (1962) 650; 
  T.~D.~Lee and M.~Nauenberg,
  Phys.\ Rev.\  {\bf 133} (1964) B1549.
  
  \bibitem{Passarino:1978jh}
  G.~Passarino and M.~J.~G.~Veltman,
  Nucl.\ Phys.\ B {\bf 160} (1979) 151.
  
\bibitem{tHooft:1978jhc} 
  G.~'t Hooft and M.~J.~G.~Veltman,
  Nucl.\ Phys.\ B {\bf 153}, 365 (1979).
  
  \bibitem{Mertig:1990an}
  R.~Mertig, M.~Bohm and A.~Denner,
  Comput.\ Phys.\ Commun.\  {\bf 64} (1991) 345.
 
 \bibitem{vanOldenborgh:1989wn}
  G.~J.~van Oldenborgh and J.~A.~M.~Vermaseren,
  Z.\ Phys.\ C {\bf 46} (1990) 425.
 
 \bibitem{Hahn:1998yk}
  T.~Hahn and M.~P\'erez-Victoria,
  Comput.\ Phys.\ Commun.\  {\bf 118} (1999) 153.
 
 \bibitem{Mathematica}
 Wolfram Research, Inc., Mathematica, Version 11.3, Champaign, IL (2018).
  
\bibitem{Illana:2000ic} 
  J.~I.~Illana and T.~Riemann,
  Phys.\ Rev.\ D {\bf 63}, 053004 (2001).
  
\bibitem{Arganda:2004bz} 
  E.~Arganda, A.~M.~Curiel, M.~J.~Herrero and D.~Temes,
  Phys.\ Rev.\ D {\bf 71}, 035011 (2005).
  
\bibitem{Patel:2015tea} 
  H.~H.~Patel,
  Comput.\ Phys.\ Commun.\  {\bf 197}, 276 (2015).

\end{thebibliography}
\end{document}